\documentclass{WileyMSP-template}
\usepackage[font=footnotesize,labelfont=bf]{caption}
\usepackage{graphicx}   
\usepackage{comment} 
\usepackage{amssymb,amsmath, mathrsfs,mathtools}
\usepackage{setspace}
\usepackage{tabularray}
\usepackage{longtable}
\usepackage{multirow}
\usepackage{array}
\usepackage{bigints}
\usepackage{xurl}
\usepackage{lineno}
\usepackage{threeparttable}
\usepackage[sort&compress,numbers]{natbib}
\usepackage[colorlinks,allcolors=blue]{hyperref}

\begin{document}
\pagestyle{fancy}
\rhead{\includegraphics[width=2.5cm]{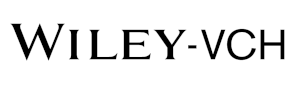}}

\title{Mechanistic Modeling of Continuous Lyophilization for Biopharmaceutical Manufacturing}

\maketitle


\author{Prakitr Srisuma}
\author{Gang Chen}
\author{Richard D. Braatz*}


\begin{affiliations}
P. Srisuma, G. Chen, R. D. Braatz\\
Massachusetts Institute of Technology, Cambridge, MA 02139, USA\\
Email: braatz@mit.edu
\end{affiliations}




\justifying
\begin{abstract}

\noindent Lyophilization (aka freeze drying) is a typical process in (bio)pharmaceutical manufacturing used for improving the stability of various drug products, including its recent applications to mRNA vaccines. While extensive efforts have been dedicated to shifting the (bio)pharmaceutical industry toward continuous manufacturing, the majority of industrial-scale lyophilization is still being operated in batch mode. This article presents the first mechanistic model for a complete continuous lyophilization process, which comprehensively incorporates and describes key transport phenomena in all three steps of lyophilization, namely freezing, primary drying, and secondary drying. The proposed model considers the state-of-the-art lyophilization technology, in which vials are suspended and move continuously through the process. The validated model can accurately predict the evolution of critical process parameters, including the product temperature, ice/water fraction, sublimation front position, and concentration of bound water, for the entire process. Several applications related to model-based process design and optimization of continuous lyophilization are also demonstrated. The final model is made available as an open-source software package that can be leveraged for guiding the design and development of future continuous lyophilization processes.

\end{abstract}


\section{Introduction} \label{ch2-sec:Intro}
Lyophilization, also known as freeze drying, is a low-temperature, low-pressure dehydration process used for improving the stability of various drug products in the biopharmaceutical industry  \cite{Fissore2018Review}. By removing the liquid component (usually water) from the product, the final lyophilized product becomes more stable, hence longer shelf life. One of the most recent applications of lyophilization is to provide long-term stability for the mRNA COVID-19 vaccines at room temperature \cite{Muramatsu2022mRNA,Meulewaeter2023mRNA}. This advancement eliminates the need for an ultra-cold supply chain for maintaining the stability of mRNA, which helps facilitate the storage and distribution of mRNA drug product across the world Advancement in lyophilization technology could therefore play a crucial role in the future of mRNA manufacturing and the biopharmaceutical industry in general.

A typical lyophilization process consists of three main steps, namely (1) freezing, (2) primary drying, and (3) secondary drying. During freezing, the product and water in a vial are cooled such that most of the liquid (free water) is frozen, with the remaining fraction (bound water) retaining its liquid state adsorbed to the solid material between the ice crystals \cite{Fissore2015Review}. In primary drying, the free water in the form of ice crystals is removed via sublimation under vacuum. Finally, in secondary drying, the product is heated further such that the bound water can be removed via desorption.

Although the current trends in the biopharmaceutical industry largely focus on the adoption of continuous manufacturing, the majority of production-scale lyophilization is still being operated in batch mode \cite{Meyer2015SpinFreezing,Pisano2019ReviewContLyo}, with most research efforts dedicated to process optimization, monitoring, and control to ensure that the final product quality meets the regulations \cite{Capozzi2019ContLyo_SuspendedVials,Pisano2019ReviewContLyo}. Conventional lyophilization of unit doses usually entails a batch of vials containing drug products situated on the cooling/heating shelf where the entire lyophilization process occurs. A number of continuous lyophilization concepts for (bio)pharmaceutical products have been proposed, including spray freeze drying \cite{Ishwarya2015SprayFD} and spin freezing \cite{Meyer2015SpinFreezing} (see a comprehensive review of continuous lyophilization technologies in \cite{Pisano2019ReviewContLyo}). A novel lyophilization technology has been recently proposed by \cite{Capozzi2019ContLyo_SuspendedVials}, in which vials are suspended and continuously move along the process without any complicated motions (e.g., as in spin or spray freeze drying), allowing the product quality control to be done more conveniently and rigorously. 

Mathematical models have been widely used to assist the design, optimization, and control of lyophilization processes. Since the key phenomena in lyophilization are underpinned by heat and mass transfer theories, process models for lyophilization are mostly physics-based rather than data-driven. Mechanistic models for batch/conventional lyophilization have been extensively discussed in the literature for many decades, e.g., for freezing \cite{Hottot2006Freezing,Nakagawa2007Freezing,Arsiccio2017CrystalSize,Deck2022FreezingLumped,Deck2024Freezing2D}, primary drying \cite{Litchfield1979Model,Liapis1994Original,Sadikoglu1997Modeling,Sheehan1998Modeling,Pikal2005Model,Hottot2006Freezing,Veraldi2008SimBatchModels,Pisano2010control,Chen2015FEMmodel,Fissore2015Review,Bano2020LumpedDrying}, and secondary drying \cite{Litchfield1979Model,Liapis1994Original,Sadikoglu1997Modeling,Sheehan1998Modeling,Pikal2005Model,Fissore2011SecDryingMonitor,Fissore2015Review,Sahni2017Simplified,Yoon2021Sec0D1D3D}. Only a few models for continuous lyophilization are available. For example, a model for the primary drying step in continuous drying of spin frozen vials was proposed by \cite{Bockstal2017ContLyo_ModelPrimary}. A model for the freezing step was also developed for spin freezing \cite{Nuytten2021SpinFreezing}. For spray freeze drying, \cite{Sebastiao2019SprayFD} developed a detailed model that can predict the temperature of droplets during the cooling and freezing phases. Nevertheless, there is no model for the suspended-vial configuration. Besides, none of the published models for continuous lyophilization considers a complete lyophilization cycle (all three steps), which is critical for optimization and control of the entire continuous operation.  

This article presents the first mechanistic model for continuous lyophilization of suspended vials. The model is developed to capture important transport phenomena in the process, including cooling and stochastic/controlled ice nucleation during the freezing step, sublimation of ice crystals during the primary drying step, and desorption of bound water during the secondary drying step. The proposed model is validated and employed to predict the evolution of critical process parameters for the entire lyophilization process, which is subsequently demonstrated for model-based design and optimization. 

This article is organized as follows. Section \ref{ch2-sec:ProcessDescription} gives an overview of conventional (batch) and continuous lyophilization technologies. Section \ref{ch2-sec:Model} extensively describes the development of our mechanistic model, supported by a comprehensive review on a variety of modeling strategies used in the literature. Section \ref{ch2-sec:Numeric} discussed the numerical methods required for solving and simulating the model equations efficiently. Section \ref{ch2-sec:Results} details the model validation and showcases various applications of the validated model. Finally, Section \ref{ch2-sec:Conclusion} summarizes the study and briefly discusses possible future work.


\section{Process Description} \label{ch2-sec:ProcessDescription}
This work considers lyophilization of unit doses, in which the product is introduced into a vial prior to being lyophilized. This type of lyophilization, compared to lyophilization of bulk material, is more preferable in (bio)pharmaceutical applications due to its accurate dosage and better control of sterility \cite{Pisano2019ReviewContLyo}. Therefore, results and discussion presented in this article are based entirely on lyophilization of unit doses.

\subsection{Batch lyophilization}
\begin{figure}[ht!]
\centering
    \includegraphics[scale=.5]{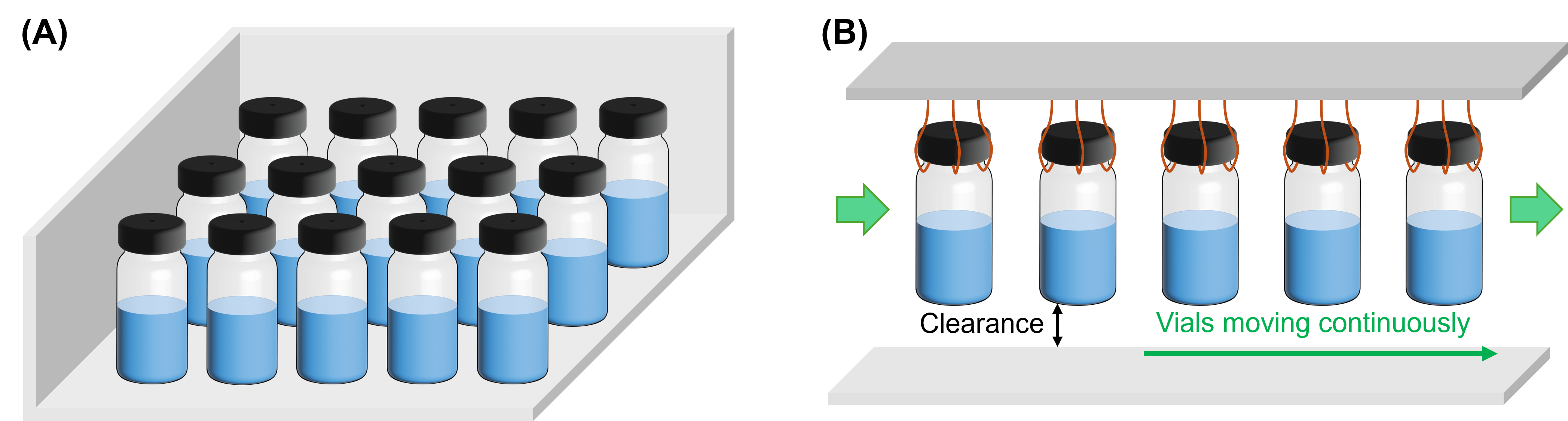}
    \caption {(A) Conventional batch lyophilization of unit doses. A number of vials are placed on the cooling/heating shelf. (B) Continuous lyophilization of suspended vials. A number of vials are suspended and move continuously through the lyophilizer.} 
    \label{fig:Vials}      
\end{figure}

The majority of lyophilization processes in the (bio)pharmaceutical industry have been operated in batch mode for many decades. The conventional configuration typically comprises a large number of vials located on the cooling/heating shelf (Figure \ref{fig:Vials}A). The bottom shelf, whose temperature can be manipulated, is used to cool the vials during the freezing step and heat the vials during the drying steps. In this case, the freezing and drying processes take place in the same space but at different times. 

Several drawbacks associated with batch lyophilization are extensively discussed in \cite{Capozzi2019ContLyo_SuspendedVials,Pisano2019ReviewContLyo}. Typical issues of any batch process include batch-to-batch variability, quality control, process downtime, and operational flexibility. Another disadvantage that is specific to the configuration illustrated in Figure \ref{fig:Vials}A is non-uniformity in heat transfer. For example, vials located on the side and corner of the shelf are typically affected by thermal radiation from the chamber walls, whereas the effect is almost negligible for center vials. Variability in heat transfer conditions could result in different crystal structures and drying times, hence variation in the final product quality. Consequently, process control and optimization are not straightforward. These known issues motivate the development of a continuous lyophilization concept that can be practically implemented in an industrial scale.

\subsection{Continuous lyophilization}
A comprehensive review of continuous lyophilization technologies is given in \cite{Pisano2019ReviewContLyo}. To summarize, the current continuous lyophilization technologies can be divided into four main categories. The first category employs equipment that continuously moves a bulk product through the process, e.g., a conveyor \cite{Gottfried1973ContLyoPatent} or a revolving plate with holes \cite{Bruttini1993ContLyoPatent}. Another type of technology relies on the concept of spray freeze drying, in which a liquid product is sprayed through nozzles to create small liquid droplets with a high surface-to-volume ratio, greatly improving heat and mass transfer in the process \cite{Ishwarya2015SprayFD,Sebastiao2019SprayFD}. Both aforementioned approaches have been proposed for lyophilization of bulk material. For lyophilization of unit doses, a well-known technique uses spin freezing to accelerate the freezing process and improve the uniformity of heat transfer in a product \cite{Meyer2015SpinFreezing}. 

Finally, the current state-of-the-art continuous lyophilization technology for unit doses has been recently proposed by \cite{Capozzi2019ContLyo_SuspendedVials}. This technology employs the suspended-vial configuration as illustrated in Figure \ref{fig:Vials}B, where all vials are suspended and continuously move through the freezing and drying chambers, without any contact between the vials and shelf. Unlike conventional lyophilizers, the freezing and drying processes occur in different locations/chambers of the equipment. This suspended-vial technology has several benefits over the existing ones. First, heat transfer in the equipment is uniform because every vial experiences the exact same heat transfer conditions along the way. In addition, there is a dedicated chamber for controlling the ice nucleation process to help reduce variation in the crystal structure caused by the stochastic nature of ice nucleation. Second, the process does not involve any complicated motions (e.g., high speed flow in spray freeze drying), resulting in a simpler design and more reliable quality control. Third, there is no contact between the vials and cooling/heating shelf, minimizing the risk of having fine particles that could lead to contamination, which is crucial in (bio)pharmaceutical applications. Finally, with the automatic filling and load-lock systems, the process is fully continuous. 

The mechanistic model developed in this work is based on this suspended-vial configuration as illustrated in Figure \ref{fig:Vials}B, the current state-of-the-art continuous lyophilization technology for unit doses. Detailed modeling strategies and model equations are discussed in the next section (Section \ref{ch2-sec:Model}), while information and discussion related to the equipment design can be found in \cite{Capozzi2019ContLyo_SuspendedVials}.  

\section{Mechanistic Modeling} \label{ch2-sec:Model}
This section details the development of a mechanistic model for continuous lyophilization of suspended vials (Figure \ref{fig:Vials}B). The section starts by discussing existing models and modeling strategies in the literature, and then derives the models for all three steps of lyophilization, namely (1) freezing, (2) primary drying, and (3) secondary drying.

\subsection{Review of the existing models and modeling strategies} \label{ch2-sec:Strategies}
Mechanistic modeling for conventional/batch lyophilization has been studied for many decades, with various models that have been proposed and used in both academia and industry. Despite the fact that the models are for batch processes, modeling strategies underpinning those models can be used as a basis for the development of a continuous lyophilization model for suspended vials. While some past reviews briefly discuss the literature on mechanistic modeling of lyophilization \cite{Veraldi2008SimBatchModels,Fissore2015Review}, no article systematically summarizes the available modeling strategies and comprehensively discusses pros and cons of each model variation. Hence, this section first provides a systematic overview of the mechanistic modeling strategies for lyophilization, discusses pros and cons of each model variation in detail, and finally concludes with the optimal modeling strategies used in this work. The published models for continuous lyophilization mentioned in Section \ref{ch2-sec:Intro} are based on the spin and spray freeze drying, which are completely different from the suspended-vial configuration, and so those modeling strategies are not discussed here.

\begin{figure}[ht!]
\centering
    \includegraphics[scale=.5]{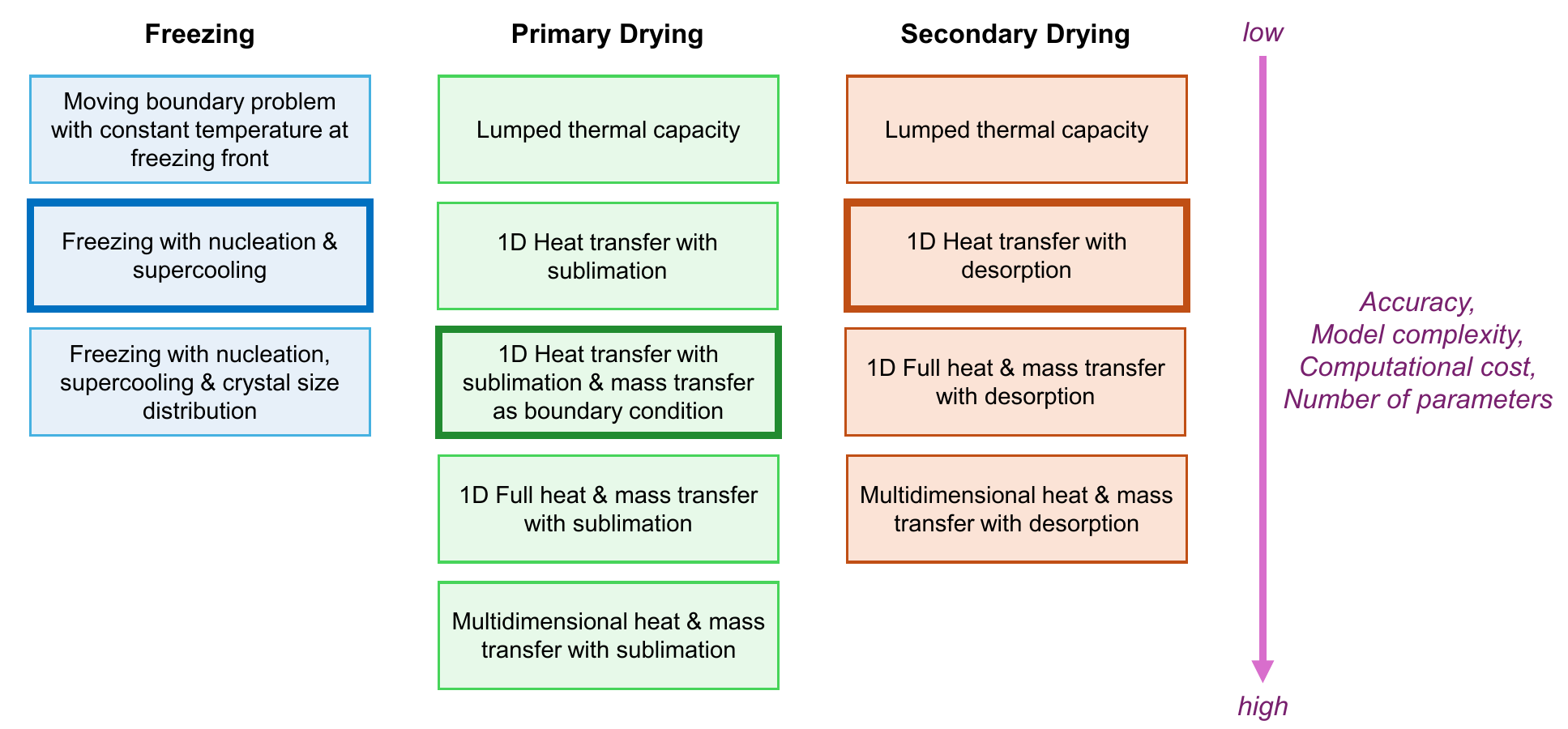}
    \caption {Modeling strategies for the freezing, primary drying, and secondary drying steps in lyophilization. The strategies used in this article are highlighted.} 
    \label{fig:ModelingStrategies}      
\end{figure}

Figure~\ref{fig:ModelingStrategies} summarizes various modeling strategies for the key steps of lyophilization, namely freezing, primary drying, and secondary drying. Published models mostly focus on primary drying because this step is recognized as the most time-consuming 
and expensive step, thus the first target for process optimization and improvement. A variety of modeling strategies are available, ranging from the simplest lumped capacity model to the high-fidelity multidimensional model. Models for secondary drying have also been well developed because the final product quality (i.e., moisture content) is governed by this step. The mechanistic modeling of freezing receives less attention in the lyophilization literature due to its complicated behavior, including supercooling and stochastic ice nucleation. As a result, the number of available models for freezing is more limited compared to that of the drying models. 

\subsubsection{Modeling strategies for the freezing step} \label{ch2-sec:strategy_freezing}
Modeling the freezing step is most complicated among the three steps of lyophilization; the blue column in Figure \ref{fig:ModelingStrategies} highlights notable modeling strategies for this step. Due to the limited number of freezing models, we expand our literature search beyond lyophilization to get more complete insights. A conventional technique for modeling the freezing process of pure substance is to apply the concept of a moving boundary problem (aka Stefan problem) where the temperature at the freezing front (solid-liquid interface) is assumed to be constant at the freezing point \cite{Mills1995HeatTransfer}. Nevertheless, this strategy has not been commonly used in the lyophilization community because it is widely known that the actual freezing process entails supercooling and stochastic ice nucleation, and so the temperature of a product usually goes below the freezing point \cite{Nakagawa2007Freezing,Deck2022FreezingLumped}. 

A more common but more complicated strategy is to consider the supercooling period and incorporate the heat transfer associated with ice nucleation as done in \cite{Hottot2006Freezing}. Recently, \cite{Deck2022FreezingLumped} introduced a state-of-the-art freezing model that incorporates the stochastic nature of ice nucleation \cite{Colucci2020CrystalSize} and considers heat transfer among multiple vials in batch lyophilization using a lumped capacity model. The same authors subsequently proposed a model for two-dimensional freezing when thermal gradients are significant, e.g., freezing of a product in a large vessel \cite{Deck2024Freezing2D}. In Appendix \ref{app:A}, we provide a detailed analysis using the relevant dimensionless group to compare between the lumped capacity model and model with thermal gradients and show that the lumped capacity model is sufficiently accurate for lyophilization of unit doses in general.

The most complicated part of freezing is to predict the crystal size or pore size distribution, which directly affects the solid structure and mass transfer resistance during sublimation in primary drying. While some recent studies have built mechanistic models with the goal of predicting pore size distribution \cite{Arsiccio2017CrystalSize,Colucci2020CrystalSize}, how to best model the specific complex interacting phenomena remains an open research question.

\subsubsection{Modeling strategies for the primary drying step} \label{ch2-sec:strategy_primary}
Many models for primary drying are available in the literature (see the green column in Figure \ref{fig:ModelingStrategies}). The simplest strategy is to use a lumped capacity model \cite{Bano2020LumpedDrying}, which does not capture the effects of spatial gradients in the system. In both batch and suspended-vial-based continuous lyophilization, vials are heated from the bottom shelf, and hence the temperature gradient in the vertical direction is significant. A more common strategy is to model the drying process in one dimension (1D, vertical direction) assuming that the process is controlled by heat transfer only \cite{Dyer1968HeatTransferLimit,Jafar2003HeatTransferLimit,Hottot2006Freezing,Srisuma2023Analytical}. By omitting mass transfer, the resulting model is simple and has only a few parameters, allowing the analytical solutions to be derived \cite{Srisuma2023Analytical}, and so the model can be implemented easily. However, as sublimation is a simultaneous heat and mass transfer process, the aforementioned models become inaccurate when the process is controlled by mass transfer. 

An approach to incorporating mass transfer into the model while avoiding any unnecessary complexity is to rely on the fact that sublimation occurs at the sublimation front/interface between the frozen and dried regions, and so mass transfer can be included as a boundary condition instead of writing a full continuity equation (e.g., see the simplified model of \cite{Veraldi2008SimBatchModels}). In this case, the driving force for mass transfer is dependent on the saturation (equilibrium) pressure at the sublimation interface, which is a function of temperature. Hence, the model should be able to accurately predict the spatial variation of the product temperature, including the interface temperature, and hence modeling the heat transfer in the vertical direction is needed. Moreover, one of the key considerations during the drying steps is to ensure that the product temperature does not exceed the upper limit (e.g., collapse temperature and glass transition temperature) \cite{Fissore2018Review}, and thus spatially distributed temperature data are valuable for accurately determining the maximum temperature in the product. 

There are models that consider detailed heat and mass transfer by including full energy and continuity equations, both in 1D \cite{Liapis1994Original,Sadikoglu1997Modeling} and in higher dimensions \cite{Sheehan1998Modeling}. Implementing such high-fidelity models for practical applications, e.g., state estimation and control, is not easy due to their high complexity, computation time, and number of parameters \cite{Veraldi2008SimBatchModels,Fissore2015Review}. Besides, the accuracy of the complex models for primary drying is not significantly different from 1D or simplified models in most cases \cite{Fissore2015Review}. Consequently, these high-fidelity models are not commonly used. 

Note that, in primary drying, the water vapor is removed via both sublimation (frozen region) and desorption (dried region). Nevertheless, it is widely acknowledged that the amount of water vapor removed via sublimation is much higher than that of desorption, and so the effect of desorption can be omitted \cite{Sadikoglu1997Modeling,Sheehan1998Modeling,Veraldi2008SimBatchModels,Fissore2015Review}. 

\subsubsection{Modeling strategies for the secondary drying step} \label{ch2-sec:strategy_secondary}
Secondary drying is similar to primary drying, with the liquid mainly removed via desorption instead of sublimation. Several modeling strategies are available, ranging from the simplest lumped capacity model \cite{Fissore2018Review} to the high-fidelity model considering multidimensional heat and mass transfer in detail \cite{Sheehan1998Modeling}. In \cite{Yoon2021Sec0D1D3D}, it was shown that the accuracy of 1D modeling is comparable to that of multidimensional modeling while being much less complicated, whereas a lumped capacity model could produce significant errors. Therefore, 1D modeling is preferable. Nevertheless, typical measurement techniques such as Karl Fischer titration measure the total amount of bound water, not the spatial variation. Consequently, a lumped capacity model is sometimes considered acceptable.

In primary drying, the drying rate is mainly governed by the rate of sublimation, which makes the modeling typically more straightforward than for secondary drying. In secondary drying, the two important mechanisms are (1) vapor phase transport through the porous dried product and (2) desorption of bound water from the solid surface of the dried product. Some previous models for secondary drying consider both phenomena \cite{Litchfield1979Model,Liapis1994Original,Sadikoglu1997Modeling,Sheehan1998Modeling}, whereas the more recently published models tend to consider only the desorption part \cite{Fissore2011SecDryingMonitor,Fissore2015Review,Sahni2017Simplified,Yoon2021Sec0D1D3D,Srisuma2024Obs}. Simulation results in these studies show that the model prediction is highly accurate even when the vapor phase transport is omitted, suggesting that desorption is the actual limiting step. In \cite{Pikal1990SecDrying_Kinetics}, a detailed experimental study was conducted, concluding that desorption is the rate-limiting step for mass transfer in secondary drying. Evidence from both simulation and experiment suggest that only the desorption part is necessary. 

Although it is evident from both simulation and experiment in the literature that desorption is the limiting step, there is no systematic analysis from the theoretical perspective of transport phenomena. In Appendix \ref{app:A}, we provide a simple and systematic way of analyzing the effect/contribution of each transport process via scale analysis, which can be used to justify the contribution of each transport process for different systems/conditions other than those considered in this work and in the literature.

\subsubsection{Optimal modeling strategies}
High-quality mechanistic modeling should balance between the model accuracy and complexity such that the resulting model can provide reliable results and be practically implemented for different purposes with ease, e.g., optimization, state estimation, and model-based control. High-fidelity models should have the best accuracy in theory, but these models typically entail a large number of equations and parameters. Simulating these model equations in real time could be challenging. Besides, incomplete knowledge of those parameters and associated uncertainty could negatively impact the model accuracy instead of improving it. On the other hand, low-fidelity models are much simpler to implement, but the model prediction might not be sufficiently accurate. 

Our goal is to select modeling strategies that can capture the critical phenomena in lyophilization and produce sufficiently accurate results while keeping the model complexity, computational cost, and number of parameters at a minimum. The resulting models should be sufficiently accurate and efficient to be used for general process design (e.g., input/output design, heat/material balance, selecting operating conditions), process optimization, state estimation, and real-time model-based control (e.g., model predictive control). For detailed process and equipment design, a high-fidelity model is needed, which is beyond the scope of this work.

From the discussion in Sections \ref{ch2-sec:strategy_freezing}--\ref{ch2-sec:strategy_secondary} and Appendix \ref{app:A}, the selected modeling strategies are highlighted in Figure \ref{fig:ModelingStrategies}. The freezing model relies on the lumped capacity method that considers nucleation and supercooling. The primary drying model simulates heat transfer in 1D (vertical direction) with mass transfer incorporated as a boundary condition at the sublimation front. Finally, the secondary drying model captures heat transfer in 1D (vertical direction) and desorption of bound water. The models for continuous lyophilization developed in the subsequent sections are based on these strategies, with proper modifications for the suspended-vial configuration. Sections \ref{ch2-sec:ModelFreezing}, \ref{ch2-sec:Model1stDrying}, and \ref{ch2-sec:Model2ndDrying} describe the models for freezing, primary drying, and secondary drying, respectively. Additionally, to supplement the mechanistic understanding of the proposed models, Section \ref{ch2-sec:heattransfer} specifically discusses modeling strategies and theories underpinning convection and thermal radiation, the key heat transfer mechanisms in suspended vials. 

\begin{figure}[ht!]
\centering
    \includegraphics[scale=1]{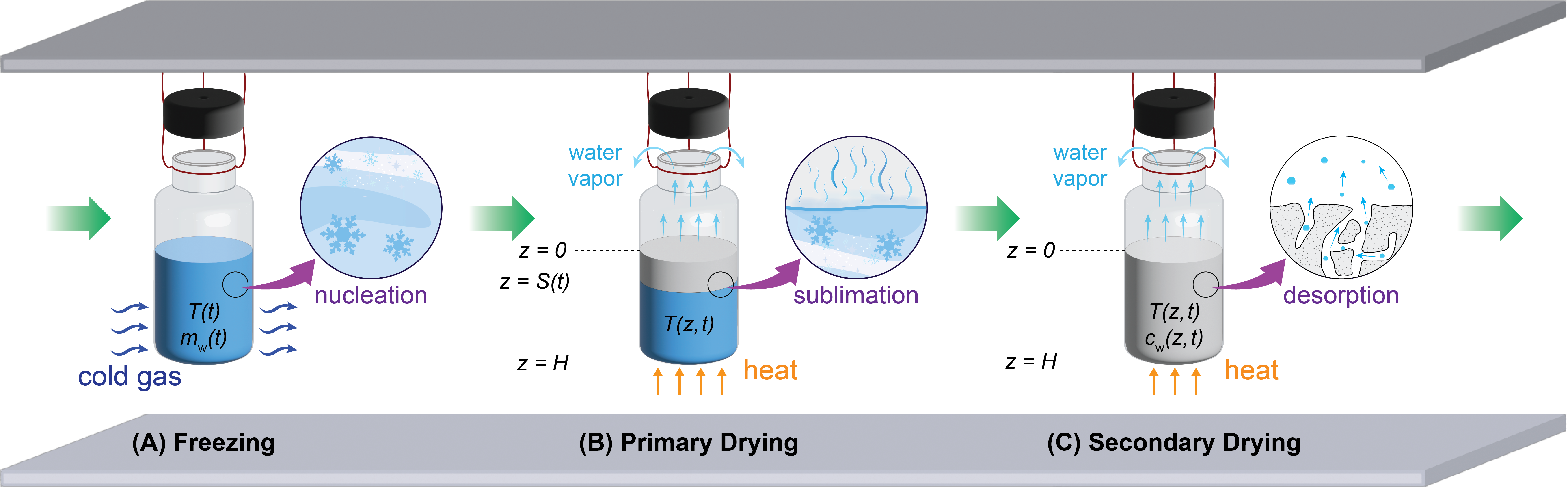}
    \caption {Schematic diagram showing the mechanistic modeling of continuous lyophilization via suspended vials for (A) freezing, (B) primary drying, and (C) secondary drying. } 
    \label{fig:Schematic}      
\end{figure}

\subsection{Model for freezing} \label{ch2-sec:ModelFreezing}
For the design proposed by \cite{Capozzi2019ContLyo_SuspendedVials}, suspended vials are cooled by using the cryogenic gas at a controlled temperature and flow rate. Besides, a dedicated chamber is added to control the nucleation temperature using the vacuum-induced surface freezing (VISF) technology \cite{Kramer2002VISF}. With this setup, all surfaces of a vial experience a similar heat transfer condition, which is different from batch lyophilization where only the bottom surface is cooled by the cooling shelf.

Our model (Figure \ref{fig:Schematic}A) for freezing of suspended vials is divided into five main steps: (1) preconditioning, (2) VISF, (3) nucleation, (4) solidification or ice formation, and (5) final cooling. 

\subsubsection{Preconditioning}\label{ch2-sec:precond}
In this model, the preconditioning step starts at $t_0$ and completes at $t_\textrm{f1}$. Preconditioning entails cooling the product in a vial such that the product temperature is uniform at the target value below the freezing point (supercooling). In terms of modeling, this step is the simplest because there is no phase change and nucleation, and so only sensible heat is important. 

The energy balance equation for the liquid solution in a vial is
\begin{equation} \label{ch2-eq:energy_precond}
    (m_\textrm{s}C_{p,\textrm{s}}+m_\textrm{w}C_{p,\textrm{w}})\frac{dT}{dt} = Q_\textrm{s1} + Q_\textrm{s2} + Q_\textrm{s3},
\end{equation}
where $T(t)$ is the temperature, $t$ is time, $m$ is the total mass, $C_p$ is the specific heat capacity (per mass), $Q$ is the total heat transfer between the vial surface and environment, the subscripts `w' and `s' refer to the water (solvent) and solid (solute) phases, and the subscripts `s1', `s2' and `s3' denote the top, bottom, and side  of the vial, respectively. The initial conditions for Equation \eqref{ch2-eq:energy_precond} are
\begin{gather}
    T(t_0) = T_0,
\end{gather}
where $T_0$ is the initial product temperature. During the freezing step in general, the amount of water changes with time due to phase transition, and so we denote $m_\textrm{w,0}$ as the initial mass of water at $t_0$. On the other hand, the amount of solid does not change, meaning that $m_\textrm{s}$ is a constant. However, during the preconditioning step, there is no phase transition at all, so
\begin{equation}
    m_\textrm{w}(t_0\leq t \leq t_\textrm{f1}) = m_\textrm{w,0}.
\end{equation}
In some cases, experimental data are reported as the total liquid volume $V_\textrm{l}$ and mass fraction of a solute $x_\textrm{s}$, where the subscript `l' denotes the liquid phase (solvent + solute). In such cases, $m_\textrm{s}$ and $m_\textrm{w,0}$ can be calculated from $V_\textrm{l}$ and the densities $\rho_\textrm{s}$, $\rho_\textrm{w}$ (see Appendix \ref{app:B} for the relevant equations).

The next step is to define the expressions for $Q_\textrm{s1},Q_\textrm{s2}$, and $Q_\textrm{s3}$, which could vary among systems. Here, we start by writing general expressions and then simplify them to match our system of interest. In suspended vials, there exists thermal radiation between the vials, chamber walls, and heating/cooling shelves. Therefore,
\begin{gather}
    Q_\textrm{s1} = h_\textrm{s1}A_z(T_\textrm{g}-T) + \sigma A_z \mathcal{F}_\textrm{s1} (T_\textrm{u}^4-T^4) , \label{ch2-eq:heat_top0} \\
    Q_\textrm{s2} = h_\textrm{s2}A_z(T_\textrm{g}-T) + \sigma A_z \mathcal{F}_\textrm{s2} (T_\textrm{b}^4-T^4), \label{ch2-eq:heat_bottom0} \\ 
    Q_\textrm{s3} = h_\textrm{s3}A_r(T_\textrm{g}-T) + \sigma A_r \mathcal{F}_\textrm{s3} (T_\textrm{c}^4-T^4), \label{ch2-eq:heat_side}   
\end{gather}
where $h$ is the heat transfer coefficient, $A_z$ is the cross sectional area of the product, $A_r$ is the side surface area of the product, $\sigma$ is the Stefan-Boltzmann constant, $\mathcal{F}$ is the transfer factor (see the definition in Section \ref{ch2-sec:heattransfer}), and the subscripts `u', `b', `c', and `g' denote the upper surface of the chamber, bottom shelf, chamber walls, and cold gas, respectively. In this work, the vial/product is modeled as a cylinder of diameter $d$, and hence the cross sectional and surface area can be calculated from the given volume and diameter. In lyophilization, the bottom shelf and gas temperatures ($T_\textrm{b}$ and $T_\textrm{g}$) typically vary with time and are specified by the freezing/drying protocol. The upper surface and wall temperatures ($T_\textrm{u}$ and $T_\textrm{c}$), however, are not usually measured and thus can be estimated from data \cite{Srisuma2024Rad}. In this article, we assume that both temperatures are approximately constant as reported by \cite{Gan2005WallTemp,Veraldi2008SimBatchModels}. In addition, $T_\textrm{u} \approx T_\textrm{c}$ due to no additional heat source/sink at these locations. Nevertheless, our model is not restricted to these assumptions; i.e., temperature-dependent data can be fed to the model (if available). 

Next, Equations \eqref{ch2-eq:heat_top0}--\eqref{ch2-eq:heat_side} are simplified further to match the experimental setup in \cite{Capozzi2019ContLyo_SuspendedVials}, whose data are used for our model validation. In \cite{Capozzi2019ContLyo_SuspendedVials}, the gas inside the chamber was cooled by the bottom shelf, and this cold gas was subsequently used to cool the vials, hence natural convection. In this setup, the upper surface and local gas in that area are not cooled, so the gas temperature is assumed to be equal to the surface temperature $T_\textrm{u}$. Consequently, the two terms can be combined and written in the form of Newton's law of cooling, with $h_\textrm{s1}$ combining the effects of both thermal radiation and natural convection, which can be done by linearizing the fourth-order term of the radiation part (see more details in Section 1.3.2 of \cite{Mills1995HeatTransfer} and Section \ref{ch2-sec:heattransfer}). Similarly, the gas at the bottom surface is cooled by the bottom shelf, so we assume that both gas and bottom shelf have the same temperature $T_\textrm{g}$, with $h_\textrm{s2}$ combining both radiation and convection. As a result, $Q_\textrm{s1}$ and $Q_\textrm{s2}$ are
\begin{gather}
    Q_\textrm{s1} = h_\textrm{s1}A_z(T_\textrm{u}-T), \label{ch2-eq:heat_top} \\
    Q_\textrm{s2} = h_\textrm{s2}A_z(T_\textrm{g}-T). \label{ch2-eq:heat_bottom}  
\end{gather}
At the side surfaces, there are natural convection from the cold gas at $T_\textrm{g}$ and thermal radiation from the chamber wall $T_\textrm{c}$, and so no simplification is needed for Equation \eqref{ch2-eq:heat_side}.

Note that the expressions for $Q_\textrm{s1},Q_\textrm{s2}$, and $Q_\textrm{s3}$ are dependent on the system. Another design proposed by \cite{Capozzi2019ContLyo_SuspendedVials} is to flush the cryogenic gas directly into the chamber. In that case, the dominant heat transfer mode is forced convection, which is much stronger than thermal radiation in this low temperature region. Thus, the effect of thermal radiation might be omitted, simplifying Equations \eqref{ch2-eq:heat_top0}--\eqref{ch2-eq:heat_side} even further.

\subsubsection{Vacuum-induced surface freezing} \label{ch2-sec:VISF}
At the end of the preconditioning step, VISF is initiated at $t_\textrm{f1}$ and proceeds until its completion at $t_\textrm{f2}$. The key idea of VISF is to reduce the total pressure to evaporate a small amount of liquid solution from the product, abruptly decreasing the temperature and promoting nucleation. An example of available models for VISF can be found in \cite{Pisano2017VISF}, which relies on the concept of the condensing/evaporating efficiency. A similar approach is used in this work, but the model is derived from the fundamental of heat and mass transfer during evaporation, e.g., as described in \cite{Mills1995HeatTransfer}, and thus the associated parameters are the heat and mass transfer coefficients instead of the condensing/evaporating efficiency. In addition, the stochastic nature of ice nucleation is also incorporated into the VISF model (described later in Section \ref{ch2-sec:Nucleation}), allowing for the comparison between spontaneous nucleation and VISF. First, consider the mass transfer part. The evaporation rate of water at the surface for a nonvolatile solute is
\begin{equation} \label{ch2-eq:evap_VISF}
    \frac{dm_\textrm{w}}{dt} = -h_mA_z(x_\textrm{w,sat}-x_\textrm{w,c}),
\end{equation}
where $h_m$ is the mass transfer coefficient, $x_\textrm{w,sat}$ is the mole fraction of water at the liquid-vapor interface (equilibrium), and $x_\textrm{w,c}$ is the mole fraction of water in the chamber (environment). The subscripts `sat' and `c' here denote the equilibrium condition and environment, respectively. To calculate the mass fraction of water for Equation \eqref{ch2-eq:evap_VISF}, we first assume that there are two gas/vapor components during VISF, namely (1) water and (2) nitrogen or inert gas, denoted by the subscript `in'. By assuming the ideal gas law,
\begin{gather}
    x_\textrm{w,sat} = \frac{p_\textrm{w,sat}M_\textrm{w}}{p_\textrm{w,sat}M_\textrm{w} + (p_\textrm{t} - p_\textrm{w,sat})M_\textrm{in}}, \\
    x_\textrm{w,c} = \frac{p_\textrm{w,c}M_\textrm{w}}{p_\textrm{w,c}M_\textrm{w} + (p_\textrm{t} - p_\textrm{w,c})M_\textrm{in}},     
\end{gather}
where $p_\textrm{t}$ is the total pressure, $p_\textrm{w}$ is the partial pressure of water, and $M$ is the molar mass. If the environment contains only nitrogen or inert gas, $p_\textrm{w,c} = 0$. The saturation pressure $p_\textrm{w,sat}$ is a function of temperature, that is \cite{Smith2018Thermo},
\begin{equation} \label{ch2-eq:psat_VISF}
    p_\textrm{w,sat} = 10^3\exp\!\left(\!16.3872-\frac{3885.7}{T-42.98}\right).
\end{equation}
Note that, if the VISF process is carried out properly, the amount of water that vaporizes is usually very small and thus does not significantly impact the overall thermophysical properties.

Next, consider the heat transfer part. The energy balance equation can be modified from Equation \eqref{ch2-eq:energy_precond} as
\begin{equation} \label{ch2-eq:energy_VISF}
    (m_\textrm{s}C_{p,\textrm{s}}+m_\textrm{w}C_{p,\textrm{w}})\frac{dT}{dt} = Q_\textrm{s1} + Q_\textrm{s2} + Q_\textrm{s3} + \Delta H_\textrm{vap}\frac{dm_\textrm{w}}{dt},  
\end{equation}
where the last term on the right-hand side is the amount of heat removed via evaporation and $\Delta H_\textrm{vap}$ is the heat of vaporization, which can be approximated by \cite{Smith2018Thermo}
\begin{equation} \label{ch2-eq:dHvap}
    \Delta H_\textrm{vap} = 2.257\mbox{$\times$}10^6\!\left(\frac{1-T/647.1}{1-373.15/647.1}\right)^{\!\!0.38}.
\end{equation}
In Equation \eqref{ch2-eq:energy_VISF}, $m_\textrm{w}(t)$ is time-dependent due to evaporation, which also means that the side surface area $A_r$ changes with time. 

The initial conditions for Equations \eqref{ch2-eq:evap_VISF} and \eqref{ch2-eq:energy_VISF} are the final mass of water and temperature at the end of the preconditioning step.

\subsubsection{Nucleation}  \label{ch2-sec:Nucleation}
The nucleation step starts at the end of VISF ($t_\textrm{f2}$) for controlled nucleation or at the end of the preconditioning step ($t_\textrm{f1}$) for uncontrolled nucleation, and then completes at $t_\textrm{f3}$. Our modeling strategy for the nucleation step described below is based on the state-of-the-art freezing model proposed by \cite{Deck2022FreezingLumped}. In this work, we use the term {\it nucleation} to denote the first nucleation where the temperature of supercooled liquid almost instantaneously increases to the equilibrium/freezing point, which is caused by the heat released from the fraction of liquid being frozen/solidified. On the other hand, we use the term {\it solidification} to denote the phase transition from liquid to solid (ice formation) after that first nucleation. 

First, consider the case of controlled nucleation with VISF. The energy balance during nucleation, assuming the process is instantaneous and adiabatic, is
\begin{equation} \label{ch2-eq:balance_nuc}
    (T_\textrm{f,l}-T_\textrm{n})(m_\textrm{s}C_{p,\textrm{s}}+m_\textrm{w}C_{p,\textrm{w}}) = m_\textrm{i,n}\Delta H_\textrm{fus},
\end{equation}
where $T_\textrm{f,l}$ is the freezing point of the liquid solution, i.e., the product temperature after nucleation, $T_\textrm{n}$ is the nucleation temperature, i.e., the product temperature when nucleation starts, $m_\textrm{i,n}$ is the mass of ice formed immediately after nucleation, the subscript `i' denotes the solid phase (ice), and $\Delta H_\textrm{fus}$ is the heat of fusion. With the presence of a non-volatile solute, the freezing-point depression is
\begin{equation} \label{ch2-eq:fp_nuc}
    T_\textrm{f,w} - T_\textrm{f,l} = \frac{K_\textrm{f}}{M_\textrm{s}}
    \!\left(\frac{m_\textrm{s}}{m_\textrm{w}-m_\textrm{i,n}}\right),
\end{equation}
where $T_\textrm{f,w}$ is the freezing point of pure water, $K_\textrm{f}$ is the molal freezing-point depression constant, and $M_\textrm{s}$ is the molar mass of a solute. The two unknowns $T_\textrm{f,l}$ and $m_\textrm{i,n}$ can be obtained by solving Equations \eqref{ch2-eq:balance_nuc} and \eqref{ch2-eq:fp_nuc} simultaneously.

Finally, define $t_\textrm{f3}$ as the time when the nucleation process completes. Since the nucleation process is nearly instantaneous as explained by \cite{Deck2022FreezingLumped}, $t_\textrm{f3}$ is set to $t_\textrm{f2}$, with the relations
\begin{gather}
    m_\textrm{w}(t_\textrm{f3}) =  m_\textrm{w}(t_\textrm{f2}) -  m_\textrm{i,n}, \label{ch2-eq:watermass_nuc}\\
    m_\textrm{i}(t_\textrm{f2}) = 0, \label{ch2-eq:icemass_nuc} \\
    m_\textrm{i}(t_\textrm{f3}) = m_\textrm{i,n}, \label{ch2-eq:icemass2_nuc} \\
    T(t_\textrm{f2}) = T_\textrm{n}, \label{ch2-eq:temp_nuc} \\
T(t_\textrm{f3}) = T_\textrm{f,l}. \label{ch2-eq:temp2_nuc}
\end{gather}

For the case of stochastic nucleation, the entire calculation described for VISF in Section \ref{ch2-sec:VISF} is skipped. By nature, ice nucleation is stochastic and can be interpreted as a Poisson process. The rate constant of the Poison process $\lambda$ is 
\begin{equation}
    \lambda = k_\textrm{n}(T_\textrm{f,l}-T)^{b_\textrm{n}}V_\textrm{l},
\end{equation}
where $k_\textrm{n}$ and $b_\textrm{n}$ are the nucleation kinetics parameters. Before nucleation occurs, there is no ice in the system, and hence $T_\textrm{f,l}$ can be calculated by
\begin{equation} \label{ch2-eq:fp2_nuc}
    T_\textrm{f,w} - T_\textrm{f,l} = \frac{K_\textrm{f}m_\textrm{s}}{M_\textrm{s}m_\textrm{w}}.
\end{equation}
The probability that the first nucleus is formed between time $t$ and $t+\Delta t$ is expressed by
\begin{equation} \label{ch2-eq:prob_nuc}
    P = 1 - \exp (\lambda \Delta t).
\end{equation}
The probability defined by Equation \eqref{ch2-eq:prob_nuc} is calculated cumulatively at every time step $\Delta t$ until the first nucleation occurs, which is marked as the end of the preconditioning step $t_\textrm{f1}$. Subsequently, follow the exact same procedure described by Equations \eqref{ch2-eq:balance_nuc}--\eqref{ch2-eq:temp2_nuc}, with $t_\textrm{f1}$ replacing $t_\textrm{f2}$.

\subsubsection{Solidification}
After the first nucleation completes, the solidification (aka ice formation) step starts at $t_\textrm{f3}$. For solidification, the lumped capacity model was shown to provide accurate prediction of the temperature and freezing time for typical vial sizes \cite{Deck2022FreezingLumped}; we also discuss this in the context of the Biot number in Appendix \ref{app:A}. The lumped capacity model is sufficiently accurate and highly computationally efficient, but it does not represent the correct physics of ice formation. By treating the entire solid/liquid as a lumped object during solidification, it implies that homogeneous nucleation is assumed. Nevertheless, it is commonly known that heterogeneous nucleation is a more frequently observed phenomenon during ice formation \cite{Glatz2018Nucleation}. Consequently, the solidification process generally starts from the vial surfaces, i.e., the outer and bottom surfaces in this case. To take this fact into account, a rigorous 2D heat transfer model could be considered \cite{Deck2024Freezing2D}. In such cases, the physics is incorporated more accurately, but the major drawback of 2D modeling is its high computational cost and numerical complexity.

\begin{figure}[ht!]
\centering
    \includegraphics[scale=.6]{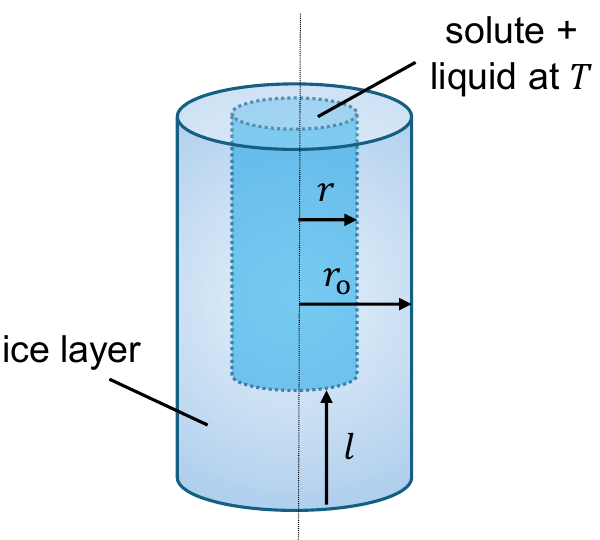}
    \caption {Schematic diagram showing the mechanistic modeling of the solidification step. For simplification, it is assumed that the liquid part retains a cylindrical shape with the same aspect ratio as the initial solution before nucleation starts.} 
    \label{fig:ice_formation}      
\end{figure}

Our strategy here is to combine the advantages of the lumped capacity and 2D models together, resulting in a model that is computationally efficient, provides accurate prediction, and represents the correct physics; we denote it as a hybrid lumped capacity model. In this case, ice formation is assumed to initiate from the side and bottom surfaces (similar to a 2D model), in which the ice layer is treated as an additional heat transfer resistance between the cold gas and unfrozen water. Additionally, heat conduction in the ice layer is assumed to be quasi steady, which is a reasonable assumption for problems associated with phase change in general as the sensible heat is much smaller than the latent heat \cite{Mills1995HeatTransfer}. Physically, this is equivalent to assuming that all the heat input is used for phase transition only. The unfrozen water and solute are treated as a lumped object as its temperature follows the equilibrium temperature. A schematic diagram showing our model for the solidification step is shown in Figure \ref{fig:ice_formation} 

The energy balance during the solidification step is 
\begin{equation} \label{ch2-eq:energy_solidification}
    (m_\textrm{s}C_{p,\textrm{s}}+m_\textrm{w}C_{p,\textrm{w}})\frac{dT}{dt} = Q_\textrm{s1} + Q_\textrm{s2} + Q_\textrm{s3} + \Delta H_\textrm{fus}\frac{dm_\textrm{i}}{dt},
\end{equation}
with $Q_\textrm{s1}$, $Q_\textrm{s2}$, and $Q_\textrm{s3}$ as defined in Equations \eqref{ch2-eq:heat_top}, \eqref{ch2-eq:heat_bottom}, and \eqref{ch2-eq:heat_side}, respectively. During the solidification step, the amount of water $m_\textrm{w}(t)$ and ice $m_\textrm{i}(t)$ varies with time, which also affects the volume and surface area of the product. The relation between $m_\textrm{w}$ and $m_\textrm{i}$ is
\begin{equation} \label{ch2-eq:mass_solidification}
m_\textrm{w} = m_\textrm{w}(t_\textrm{f3}) - m_\textrm{i}.
\end{equation} 
The temperature of the liquid phase and the solid-liquid interface follows the freezing-point depression relation 
\begin{equation} \label{ch2-eq:temp_solidification}
    T = T_\textrm{f,w} - \frac{K_\textrm{f}}{M_\textrm{s}}
    \!\left(\frac{m_\textrm{s}}{m_\textrm{w}}\right).
\end{equation}
The ice layer becomes an additional heat transfer resistance, and so the heat transfer coefficients  $h_\textrm{s2}$ and $h_\textrm{s3}$ in Equations \eqref{ch2-eq:heat_bottom} and \eqref{ch2-eq:heat_side} can be replaced by the overall heat transfer coefficients
\begin{gather}
   U_\textrm{s2} = \frac{ 1}{1/h_\textrm{s2} + l/k_\textrm{i}}, \label{ch2-eq:heat_bottom2}  \\
   U_\textrm{s3} = \frac{1}{1/h_\textrm{s3} + \frac{r_\textrm{o}\ln(r_\textrm{o} / r)}{k_\textrm{i}}} .\label{ch2-eq:heat_side2} 
\end{gather}
where $r_\textrm{o}$, $r$ and $l$ are as defined in Figure \ref{fig:ice_formation} and $k_\textrm{i}$ is the thermal conductivity of ice. For the radiation part, the linearization technique described in Section \ref{ch2-sec:precond} can be applied, and so the equation is written in the form of Newton's law of cooling. Note that the overall heat transfer coefficients in Equations \eqref{ch2-eq:heat_bottom2} and \eqref{ch2-eq:heat_side2} are defined based on the outer surface area to be consistent with the original expressions of $Q_\textrm{s2}$ and $Q_\textrm{s3}$ .

The initial conditions for $m_\textrm{w}$, $m_\textrm{i}$, and $T$ are the final conditions of the nucleation step, namely Equations \eqref{ch2-eq:watermass_nuc}, \eqref{ch2-eq:icemass2_nuc}, and \eqref{ch2-eq:temp2_nuc}. Finally, define $t_\textrm{f4}$ as the time when solidification completes. The criterion for complete solidification is defined as
\begin{equation} \label{ch2-eq:end_solidification}
    m_\textrm{i}(t_\textrm{f4}) = 0.95m_\textrm{w}(t_\textrm{f3}),
\end{equation}
where the coefficient could vary between 0.85 and 0.95 without significantly changing the final results \cite{Deck2022FreezingLumped}.

Our hybrid lumped capacity model presented in this section has the same model complexity and computational cost as those of the original lumped capacity model (e.g., in \cite{Deck2022FreezingLumped}) while describing the physics in a more accurate way as in the 2D model (e.g., in \cite{Deck2024Freezing2D}). Final simulations results from these models are almost identical for lyophilization of unit doses, and so there is no concern in terms of model accuracy.

\subsubsection{Cooling}\label{ch2-sec:cooling}
The final step is to ensure that the product temperature is at the desire value before starting the drying process. This cooling step starts at $t_\textrm{f4}$ and completes at $t_\textrm{f5}$. The energy balance is
\begin{equation} \label{ch2-eq:energy_cooling}  
    (m_\textrm{s}C_{p,\textrm{s}}+m_\textrm{w}C_{p,\textrm{w}}+m_\textrm{i}C_{p,\textrm{i}})\frac{dT}{dt} = Q_\textrm{s1} + Q_\textrm{s2} + Q_\textrm{s3},
\end{equation}
with $Q_\textrm{s1}$, $Q_\textrm{s2}$, and $Q_\textrm{s3}$ as defined in Equations \eqref{ch2-eq:heat_top}, \eqref{ch2-eq:heat_bottom}, and \eqref{ch2-eq:heat_side}, respectively. Since there is no phase change during this step, the amount of substances, volume, and surface area are all constant, following the final conditions of the solidification step. The initial condition for Equation \eqref{ch2-eq:energy_cooling} is the final temperature of the solidification step. It is generally known that, after solidification, the amount of bound water is negligible compared to that of ice, and so $m_\textrm{w}C_{p,\textrm{w}}$ could be omitted from Equation \eqref{ch2-eq:energy_cooling} without any significant error.

By consecutively simulating the models developed in Sections \ref{ch2-sec:precond} to \ref{ch2-sec:cooling}, the evolution of the product temperature, phase transition, and amount of ice/water during the freezing step can be predicted.

\subsection{Model for primary drying} \label{ch2-sec:Model1stDrying}
In conventional lyophilization, vials are placed on the heating shelf, and thus heat transfer at the bottom surface is driven by conduction (at the point of contact), convection (gas), and radiation, whereas only radiation dominates at the side and top surfaces \cite{Pikal2005Model}. For continuous lyophilization, suspended vials are heated by the below heating shelf without any contact between the vials and shelf. As a result, heat transfer at the bottom surface is driven only by thermal radiation and natural convection. 

The model for primary drying is formulated in the rectangular coordinate system with one spatial dimension ($z$) and time ($t$) (Figure \ref{fig:Schematic}B). Define the primary drying step to start at $t_0$ and complete at $t_\textrm{d1}$. If the primary drying model is simulated consecutively after the freezing step, then set $t_0 = t_\textrm{f4}$. Otherwise, $t_0$ should be set to 0 for a standalone primary drying simulation. 

The governing equations for the primary drying step consist of the (1) energy balance in the frozen region and (2) mass balance at the sublimation front. By assuming that the supplied heat is used in the frozen region only, the energy balance for the frozen region can be described by the partial differential equation (PDE)
\begin{equation} \label{ch2-eq:1st_energy}
    \rho_\textrm{f} C_{p,\textrm{f}}\dfrac{\partial T}{\partial t} = k_\textrm{f}\dfrac{\partial^2 T}{\partial z^2} + \frac{Q_\textrm{rad}}{V_\textrm{f}}, \qquad  S < z < H, 
\end{equation}
where $T(z, t)$ is the temperature, $S(t)$ is the sublimation front/interface position, $k$ is the thermal conductivity, $\rho$ is the density, $C_p$ is the heat capacity, $V$ is the volume, $H$ is the height of the product, and the subscript `f' denotes the frozen region. We refer to Appendix \ref{app:B} for the calculations of some relevant parameters. The radiative heat transfer from the sidewall $Q_\textrm{rad}$ is
\begin{equation} \label{ch2-eq:1st_rad}
    Q_\textrm{rad} = \sigma A_r \mathcal{F}_\textrm{s3} \!\left(T_\textrm{c}^4-T^4\right),
\end{equation}
where $T_\textrm{c}$ is the chamber wall temperature and $A_r=\pi dH$ is the side area of the product. We refer to Section \ref{ch2-sec:heattransfer} for the detailed derivation of Equation \eqref{ch2-eq:1st_rad}. The mass balance of water at the sublimation front gives
\begin{equation}\label{ch2-eq:1st_interface}
    \frac{dS}{dt} = \frac{N_\textrm{w}}{\rho_\textrm{f}-\rho_\textrm{e}},
\end{equation}
where $N_\textrm{w}$ is the sublimation flux and $\rho_\textrm{e}$ is the effective density of the dried region above the sublimation front. The driving force for mass transfer at the sublimation interface is \cite{Pikal2005Model,Fissore2018Review,Bano2020LumpedDrying}
\begin{equation}\label{ch2-eq:1st_flux}
    N_\textrm{w} = \frac{p_\textrm{w,sat}-p_\textrm{w,c}}{R_\textrm{p}},
\end{equation}
where $p_\textrm{w,sat}$ is the saturation/equilibrium pressure of water, $p_\textrm{w,c}$ is the partial pressure of water in the chamber (environment), and $R_\textrm{p}$ is the mass transfer resistance. The saturation pressure for sublimation is described by \cite{Bano2020LumpedDrying}
\begin{equation}\label{ch2-eq:1st_satpressure}
    p_\textrm{w,sat} =  \exp\!\left(\frac{-6139.9}{T} + 28.8912\right).
\end{equation}
The variation of the mass transfer resistance can be approximated by the empirical expression \cite{Pikal2005Model,Fissore2015Review,Bano2020LumpedDrying}
\begin{equation}\label{ch2-eq:1st_Rp}
    R_\textrm{p} = R_\textrm{p0}+ \frac{R_\textrm{p1}S}{R_\textrm{p2}+S},
\end{equation}
where $R_\textrm{p0}$, $R_\textrm{p1}$, and $R_\textrm{p2}$ are the constants to be estimated from data. Physically, $R_\textrm{p0}$ represents the resistance associated with mass convection above the product, which could include the presence of a skin layer on the cake surface and the effect of the stopper \cite{Stratta2024Stopper}, whereas $\frac{R_\textrm{p1}S}{R_\textrm{p2}+S}$ corresponds to the resistance associated with diffusion through the porous dried layer in the product. 

Modeling the heat transfer and sublimation front in one dimension, as described in Equations \eqref{ch2-eq:1st_energy} and \eqref{ch2-eq:1st_interface}, implies an assumption that the sublimation front is flat. This assumption has been widely used for conventional lyophilization as the heat flux through the bottom of the vial is much stronger than that from the sidewall, mainly due to heat conduction. In lyophilization of suspended vials, due to the absence of conductive heat transfer at the bottom surface, the sidewall heat flux becomes more comparable to the bottom heat flux. However, the bottom heat flux still dominates for two key reasons. First, the bottom shelf remains the primary heat source in the system, which directly interacts with the bottom surface of the vial through thermal radiation and convection. Second, since most vials are surrounded by others at similar temperatures, the driving force for sidewall heat transfer is significantly weaker than that for bottom heat transfer. A systematic analysis for comparing different heat transfer modes can be done using relevant dimensionless numbers, e.g., in Appendix \ref{app:A}.

The PDE represented by Equation \eqref{ch2-eq:1st_energy} requires two boundary conditions. Heat transfer at the bottom surface of the frozen product follows Newton's law of cooling,
\begin{equation}\label{ch2-eq:1st_bcbottom}
    - k_\textrm{f}\dfrac{\partial T}{\partial z} = h_\textrm{b}(T-T_\textrm{b}),  \qquad z=H, 
\end{equation}
where $T_\textrm{b}$ is the bottom shelf temperature and $h_\textrm{b}$ is the overall heat transfer coefficient that combines the effects of thermal radiation and natural convection. At the top surface, the energy balance associated with sublimation and thermal radiation from the upper surface of the chamber is
\begin{equation}\label{ch2-eq:1st_bctop}
    N_\textrm{w}\Delta H_\textrm{sub}  =  k_\textrm{f}\dfrac{\partial T}{\partial z} + \sigma\mathcal{F}_\textrm{s1} \big(T_\textrm{u}^4-T^4\big), \qquad z=S,
\end{equation}
where $\Delta H_\textrm{sub}$ is the heat of sublimation.

The initial conditions for Equations \eqref{ch2-eq:1st_energy} and \eqref{ch2-eq:1st_interface} are
\begin{align}
    T(z, t_0) &= T_0, \qquad 0\leq z \leq H, \label{ch2-eq:1st_iniT}  \\
    S(t_0) &=  0. \label{ch2-eq:1st_iniS} 
\end{align}
For consecutive simulation with the freezing step, set $T_0 = T(t_\textrm{f4})$. Otherwise, $T_0$ can be set arbitrarily for a standalone primary drying simulation.

The primary drying model is simulated until the interface position is equal to the height of the product, i.e., $S = H$, indicating that there is no frozen material left, which marks the end of the primary drying step at $t_\textrm{d1}$. In some cases, there might be an additional heating period at the end of primary drying to adjust the temperature and ensure complete sublimation before starting the secondary drying step. However, the model contains only a simple heat equation, and so it is not detailed here.

\subsection{Model for secondary drying} \label{ch2-sec:Model2ndDrying}
The model for secondary drying is formulated in the rectangular coordinate system with one spatial dimension ($z$) and time ($t$) (Figure \ref{fig:Schematic}C), which is consistent with the primary drying model. The secondary drying step is defined to start at $t_0$ and complete at $t_\textrm{d2}$. If the secondary drying model is simulated consecutively after the primary drying step, then set $t_0 = t_\textrm{d2}$. Otherwise, $t_0$ should be set to 0 for a standalone secondary drying simulation. 

The governing equations for the secondary drying step comprise the (1) energy balance in the dried region and (2) desorption kinetics. The energy balance of the dried product is 
\begin{equation} \label{ch2-eq:2nd_energy}
    \rho_\textrm{e} C_{p,\textrm{e}}\dfrac{\partial T}{\partial t} = k_\textrm{e}\dfrac{\partial^2 T}{\partial z^2}  + \rho_\textrm{d}\Delta H_\textrm{des}\dfrac{\partial c_\textrm{w}}{\partial t} + \frac{Q_\textrm{rad}}{V_\textrm{e}},  \qquad  0 \leq z \leq H,
\end{equation}
where $T(z, t)$ is the product temperature, $c_\textrm{w}(z, t)$ is the concentration of bound water (aka moisture content, residual moisture, residual water), $\rho_\textrm{d}$ is the density of the dried region (solid and vacuum), $\Delta H_\textrm{des}$ is the heat of desorption, $Q_\textrm{rad}$ is as defined in Equation \eqref{ch2-eq:1st_rad}, and the other parameters are as defined in Equation \eqref{ch2-eq:1st_energy}, with the subscript `e' denoting the effective properties considering both solid and gas in the pores. The desorption kinetics of bound water is described by
\begin{equation} \label{ch2-eq:2nd_desorption}
     \dfrac{\partial c_\textrm{w}}{\partial t} = k_\textrm{d}(c^*_\textrm{w}-c_\textrm{w}),
\end{equation}
where $c^*_\textrm{w}$ is the equilibrium concentration of bound water and $k_\textrm{d}$ is the rate constant for desorption that exhibits Arrhenius temperature dependence \cite{Liapis1994Original,Sadikoglu1997Modeling,Fissore2015Review} 
\begin{equation} \label{ch2-eq:2nd_kd}
     k_\textrm{d} = f_\textrm{a}e^{-E_\textrm{a}/RT},
\end{equation}
where $f_\textrm{a}$ is the frequency factor (aka collision frequency),  $E_\textrm{a}$ is the activation energy, and $R$ is the gas constant. The above desorption kinetics, Equation \eqref{ch2-eq:2nd_desorption}, is known as the linear driving force model, one of the simplest adsorption/desorption models that can accurately predict the dynamics of bound water and has been widely used in the literature \cite{Liapis1994Original,Sircar2000LDF,Veraldi2008SimBatchModels,Fissore2011SecDryingMonitor,Fissore2015Review}. Further simplification that is relatively common can be done by setting $c^*_\textrm{w}=0$. This simplification produces insignificant error as shown in \cite{Sadikoglu1997Modeling} and eliminates the need for equilibrium data and detailed knowledge about the solid structure \cite{Fissore2015Review}.

The governing PDE, Equation \eqref{ch2-eq:2nd_energy}, requires two boundary conditions. The bottom surface of the dried product is heated by the heating shelf, which follows Newton's law of cooling
\begin{equation} \label{ch2-eq:2nd_bctbottom}
    - k_\textrm{e}\dfrac{\partial T}{\partial z} = h_\textrm{b}(T-T_\textrm{b}),  \qquad z=H, 
\end{equation}
where the value of $h_\textrm{b}$ is approximated to have the same value as that used in primary drying, Equation \eqref{ch2-eq:1st_bcbottom}. Heat transfer at the top surface is mainly thermal radiation, resulting in the boundary condition
\begin{equation} \label{ch2-eq:2nd_bctop}
     -k_\textrm{e}\dfrac{\partial T}{\partial z} =  \sigma\mathcal{F}_\textrm{s1}  \big(T_\textrm{u}^4-T^4\big), \qquad  z=0.
\end{equation}

The initial conditions for Equations \eqref{ch2-eq:2nd_energy} and \eqref{ch2-eq:2nd_desorption} are
\begin{gather} 
    T(z, t_0) = T_0, \qquad 0\leq z \leq H, \label{ch2-eq:2nd_iniT}  \\
    c_\textrm{w}(z, t_0) = c_\textrm{w,0},  \qquad  0 \leq z \leq H. 
\end{gather}
For consecutive simulation with the primary drying step, set $T_0 = T(z, t_\textrm{d1})$. Otherwise, $T_0$ can be set arbitrarily for a standalone secondary drying simulation.

The secondary drying model should be simulated until the concentration of bound water is below the target value, denoted as $c_{\textrm{w,}\infty}$, which marks the end of the secondary drying step at $t_\textrm{d2}$. 

\subsection{Modeling heat transfer in suspended vials} \label{ch2-sec:heattransfer}
Convection and radiation are the two important modes of heat transfer in suspended vials as discussed extensively in the previous sections. During the freezing step, there could be a combination of natural/forced convection and radiation, depending on how the cryogenic gas is fed to the system. In primary and secondary drying, only natural convection and radiation are important.

Convection can be simply modeled using Newton's law of cooling. Heat transfer coefficients for convection can be estimated from correlations that entail dimensionless groups such as the Nusselt number, Prandtl number, Reynolds number (forced convection), and Grasholf number (natural convection). Nevertheless, it is more common and accurate to estimate the heat transfer coefficient from data, e.g., using the techniques suggested in \cite{Fissore2015Review}. 

Thermal radiation is significantly more complicated in terms of modeling. The theories and model equations described below are mainly based on in \cite{Mills1995HeatTransfer} and \cite{Srisuma2024Rad}; the former discusses general thermal radiation theories, while the latter provides a detailed study on the modeling of thermal radiation in lyophilization. Here we summarize only the key elements needed for modeling thermal radiation in the suspended-vial configuration; more detailed discussion can be found in the aforementioned references. Consider radiation exchange between two diffuse, gray surfaces of finite size denoted as surfaces 1 and 2, respectively. In general, the net radiant energy receiving by surface 1 can be written as 
\begin{equation} \label{ch2-eq:rad_general}
    Q_\textrm{rad} = \sigma A_1 \mathcal{F} \big(T_2^4- T_1^4\big),
\end{equation}
where $A$ is the surface area, $\mathcal{F}$ is the transfer factor, and the subscripts `1' and `2' denote surfaces 1 and 2, respectively. The transfer factor is dependent on the geometry and material properties (emissivity) of both surfaces. For any two diffuse, gray surfaces that form an enclosure,
\begin{equation} \label{ch2-eq:rad}
    Q_\textrm{rad} = \dfrac{\sigma\big(T_2^4- T_1^4\big)}{\dfrac{1-\varepsilon_1}{\varepsilon_1A_1} + \dfrac{1}{A_1F_{1-2}} + \dfrac{1-\varepsilon_2}{\varepsilon_2A_2}},
\end{equation}
where $F_{1-2}$ is the view/shape factor. If surface 1 is surrounded by surface 2 (i.e., $F_{1-2}=1$) and surface 2 is much larger than surface 1, Equation \eqref{ch2-eq:rad} can be simplified as
\begin{equation} \label{ch2-eq:rad_sim}
    Q_\textrm{rad} = \varepsilon_1\sigma A_1\big(T^4_2-T^4_1\big).
\end{equation}
In Equation \eqref{ch2-eq:rad_sim}, the transfer factor $\mathcal{F}=\varepsilon_1$. Next, consider the case where a single vial (surface 1) is surrounded by the chamber walls (surface 2). In this case, Equation \eqref{ch2-eq:rad_sim} becomes
\begin{equation} \label{ch2-eq:rad2}
    Q_\textrm{rad} = \varepsilon_\textrm{gl}\sigma A_r\big(T^4_\textrm{c}-T^4\big),
\end{equation}
where $\varepsilon_\textrm{gl}$ is the emissivity of the glass vial. Here the transfer factor $\mathcal{F} = \varepsilon_\textrm{gl}$. The amount of radiative heat for the single-vial case  represented by Equation \eqref{ch2-eq:rad2} defines the upper bound on the radiative heat that one vial can receive from the chamber walls. This is because, when there are multiple vials, the view factor $F_{1-2}$ is less than 1, and so the radiative heat is shared among the vials. As a result, the transfer factor $\mathcal{F}$ for the multiple-vial case must be less than $\varepsilon_\textrm{gl}$. Therefore, Equation \eqref{ch2-eq:rad2} needs to be modified to account for interactions between vials.

Typical lyophilization of unit doses always consists of a large number of vials, and so thermal radiation exchange exists not only between vials and chamber walls but also between those multiple vials. In batch lyophilization, vials are arranged as an array consisting of many rows, which results in significant differences in heat transfer conditions between the outer and inner vials. In such cases, the outer vials with higher temperature transfer significant heat to the inner vials while simultaneously exchanging heat with the chamber walls. A rigorous way of modeling this complicated radiation is the radiation network approach, which is comprehensively discussed and demonstrated in the context of lyophilization in \cite{Srisuma2024Rad}. In the suspended-vial configuration, however, vials are typically aligned in a few rows, e.g., a single row in the equipment proposed by \cite{Capozzi2019ContLyo_SuspendedVials}. As a result, all vials experience nearly the same heat transfer heat condition, resulting in negligible heat transfer between vials. Therefore, instead of using the radiation network approach, we modify Equation \eqref{ch2-eq:rad2} to follow Equation \eqref{ch2-eq:rad_general} as
\begin{equation} \label{ch2-eq:rad4}
    Q_\textrm{rad} = \sigma A_r\mathcal{F}\big(T^4_\textrm{c}-T^4\big),
\end{equation}
where $\mathcal{F}$ must be less than $\varepsilon_\textrm{gl}$ as described in the previous paragraph. This transfer factor $\mathcal{F}$ is best to be estimated from experimental data but can also be approximated mechanistically, e.g., using Equation \eqref{ch2-eq:rad}. In our model, the transfer factor $\mathcal{F}_\textrm{s3}$ appears in Equations \eqref{ch2-eq:heat_side}, \eqref{ch2-eq:1st_energy}, and \eqref{ch2-eq:2nd_energy}. If the chamber design and vial configuration are identical for both freezing and drying steps, the same value of $\mathcal{F}_\textrm{s3}$ can be used for all equations. A similar analysis can also be done for $\mathcal{F}_\textrm{s1}$ in Equations \eqref{ch2-eq:1st_bctop} and \eqref{ch2-eq:2nd_bctop}.

When there exists both convection and radiation, it is more convenient to linearize and rewrite the radiation part in the form of Newton's law of cooling; at the temperature of about 300 K, the error caused by linearization is about 0.1\% for the temperature difference of 20 K and 2\% for the temperature difference of 100 K (see Section 1.3.2 of \cite{Mills1995HeatTransfer}), which is tiny. The linearized equation can then be combined with the convection part, with the corresponding heat transfer coefficient taking into account of both convection and radiation, e.g., Equations \eqref{ch2-eq:heat_bottom}, \eqref{ch2-eq:1st_bcbottom}, and \eqref{ch2-eq:2nd_bctbottom}. This approximation helps simplify the equations and also facilitates the use of an overall heat transfer coefficient (e.g., Equation \eqref{ch2-eq:heat_side2}) as both convection and radiation parts are written in the form of Newton's law of cooling. 

Finally, to ensure physically reasonable heat transfer parameters in our model, we note that typical heat transfer coefficients for natural convection in air vary between 3 and 25 W/m$^2$$\cdot$K, and those for forced convection vary between 10 and 200 W/m$^2$$\cdot$K \cite{Mills1995HeatTransfer}.

\subsection{Default model parameters}
This section defines the default parameter values for the model developed in Sections \ref{ch2-sec:ModelFreezing}, \ref{ch2-sec:Model1stDrying}, and \ref{ch2-sec:Model2ndDrying}. These parameter values are either obtained from the literature or set based on the typical values used in lyophilization. This default set of parameters assumes a complete simulation in which the models for freezing (with controlled nucleation), primary drying, and secondary drying are simulated consecutively. Table \ref{ch2-tab:DefaultParameters} lists the default model parameters; parameter values different from those reported in the table are stated explicitly in each section or case study.

{\singlespacing\renewcommand{\arraystretch}{1.2}
\begin{longtable}[ht!]{llll} 
\caption{Default model parameters for a complete continuous lyophilization simulation.} 
\label{ch2-tab:DefaultParameters}\\
\hline
\textbf{Symbol} &\textbf{Value} & \textbf{Unit} & \textbf{Source} \\
\hline
\multicolumn{4}{l}{\textbf{Thermophysical properties}}  \\ 
$C_{p,\textrm{e}}$ & 2590 & J/kg$\cdot$K & \cite{Sadikoglu1997Modeling}  \\
$C_{p,\textrm{f}}$ & 2163 & J/kg$\cdot$K &  calculated (see Appendix \ref{app:B})   \\
$C_{p,\textrm{i}}$ & 2108 & J/kg$\cdot$K & \cite{Deck2024Freezing2D} \\
$C_{p,\textrm{s}}$ & 1204 & J/kg$\cdot$K &  \cite{Deck2024Freezing2D} \\
$C_{p,\textrm{w}}$ & 4.187  & J/kg$\cdot$K &  \cite{Deck2024Freezing2D}  \\
$\Delta H_\textrm{des}$ &  2.68$\times$10$^{6}$  & J/kg & \cite{Sadikoglu1997Modeling} \\ 
$\Delta H_\textrm{fus}$ & 3.34$\times$10$^{5}$ & J/kg & \cite{Smith2018Thermo} \\ 
$\Delta H_\textrm{sub}$ & 2.84$\times$10$^{6}$ & J/kg & \cite{Veraldi2008SimBatchModels} \\ 
$\Delta H_\textrm{vap}$ & see Equation \eqref{ch2-eq:dHvap} & J/kg & \cite{Smith2018Thermo} \\ 
$k_\textrm{e}$ & 0.217 & W/m$\cdot$K & \cite{Sadikoglu1997Modeling}  \\
$k_\textrm{f}$ & 2.07 & W/m$\cdot$K &  calculated (see Appendix \ref{app:B})   \\
$k_\textrm{i}$ & 2.25 & W/m$\cdot$K & \cite{Deck2024Freezing2D} \\
$k_\textrm{s}$ & 0.126 & W/m$\cdot$K &  \cite{Deck2024Freezing2D} \\
$k_\textrm{w}$ & 0.598 & W/m$\cdot$K &  \cite{Deck2024Freezing2D}  \\
$\rho_{\textrm{e}}$ & 215 & kg/m$^3$ & \cite{Sadikoglu1997Modeling}  \\
$\rho_\textrm{d}$ & 212.21 & kg/m$^3$ & \cite{Sadikoglu1997Modeling}  \\
$\rho_\textrm{f}$ & 937 & kg/m$^3$ &  calculated (see Appendix \ref{app:B})  \\
$\rho_\textrm{i}$ & 917 & kg/m$^3$ & \cite{Colucci2020CrystalSize} \\
$\rho_\textrm{s}$ & 1587.9 & kg/m$^3$ &  \cite{Colucci2020CrystalSize} \\
$\rho_\textrm{w}$ & 1000  & kg/m$^3$ &  \cite{Deck2024Freezing2D}  \\
\hline \multicolumn{4}{l}{\textbf{Operating conditions}}  \\ 
$c_\textrm{w,0}$ & 0.088 & kg water/kg solid & \cite{Capozzi2019ContLyo_SuspendedVials} \\
$c_{\textrm{w,}\infty}$ &  0.01 & kg water/kg solid & \cite{Fissore2018Review} \\
$p_\textrm{t}$ & $10^5$ (freezing) & Pa & -- \\
& $10^4$ (VISF) &  Pa & --  \\
$p_\textrm{w,c}$ & 3 & Pa & \cite{Capozzi2019ContLyo_SuspendedVials} \\
$p_\textrm{w,sat}$ & see Equations \eqref{ch2-eq:psat_VISF} and \eqref{ch2-eq:1st_satpressure} & Pa & \cite{Smith2018Thermo,Bano2020LumpedDrying} \\
$T_0$ & 298.15 & K & -- \\
$T_\textrm{b}$ & 270 (primary drying) & K & --  \\
& 295 (secondary drying) & K & --  \\
$T_\textrm{c},T_\textrm{u}$ & 273 (before VISF) & K &  -- \\
& 240 (after VISF) & K & --  \\
& 265 (primary drying) & K & --  \\
& 290 (secondary drying) & K & --  \\
$T_\textrm{f,w}$ & 273.15 & K & \cite{Deck2024Freezing2D} \\
$T_\textrm{g}$ & 268 (before VISF) & K & -- \\
& 230 (after VISF) & K & -- \\
$T_\textrm{n}$ & 268 & K & -- \\
\hline \multicolumn{4}{l}{\textbf{Heat and mass transfer}}  \\ 
$\mathcal{F}_\textrm{s1}$ & 0.8 & -- & -- \\ 
$\mathcal{F}_\textrm{s3}$ & 0.624 & -- & -- \\ 
$f_\textrm{a}$  & 1.5$\times$10$^{-3}$ & 1/s & -- \\ 
$E_\textrm{a}$ & 6500 & J/mol$\cdot$K & -- \\
$h_\textrm{b}$ & 15 & W/m$^2$$\cdot$K & -- \\
$h_\textrm{m}$ & 6.34$\times$10$^{-3}$ & kg/m$^2$$\cdot$s & \cite{Mills1995HeatTransfer} \\
$h_\textrm{s1}$ & 5 & W/m$^2$$\cdot$K & -- \\
$h_\textrm{s2}$ & 10 & W/m$^2$$\cdot$K & -- \\
$h_\textrm{s3}$ & 8 & W/m$^2$$\cdot$K & -- \\
$R_\textrm{p0}$ & 1.5$\times$10$^{4}$ & m/s &  -- \\
$R_\textrm{p1}$ & 3.0$\times$10$^{7}$ & 1/s &  -- \\
$R_\textrm{p2}$ & 10 & 1/m &  -- \\
$\sigma$ & 5.67$\times$10$^{-8}$  & W/m$^2$$\cdot$K$^4$ & -- \\ 
\hline \multicolumn{4}{l}{\textbf{Product and formulation}}  \\ 
$H$ &7.2$\times$10$^{-3}$ & m & calculated (see Appendix \ref{app:B})  \\
$m_\textrm{s}$ & 1.53$\times$10$^{-4}$ & kg & calculated (see Appendix \ref{app:B})   \\ 
$m_\textrm{w,0}$ & 2.9$\times$10$^{-3}$ & kg & calculated (see Appendix \ref{app:B}) \\ 
$V_\textrm{l}$ & 3$\times$10$^{-6}$ & m$^3$ & \cite{Capozzi2019ContLyo_SuspendedVials}  \\ 
$x_\textrm{s}$ & 0.05 & -- & \cite{Capozzi2019ContLyo_SuspendedVials} \\ 
\hline \multicolumn{4}{l}{\textbf{Vial}}  \\ 
$d$ & 0.024  & m & 10R vial  \\ 
$\epsilon_\textrm{gl}$ & 0.8  & -- & \cite{Mills1995HeatTransfer}  \\ 
\hline \multicolumn{4}{l}{\textbf{Others}}  \\ 
$K_\textrm{f}$ & 1.86  & kg$\cdot$K/mol & \cite{Chang2010Chem}  \\ 
$M_\textrm{in}$ & 0.028  & kg/mol & \cite{Chang2010Chem} \\ 
$M_\textrm{s}$ & 0.3423  & kg/mol & \cite{Deck2024Freezing2D} \\ 
$M_\textrm{w}$ & 0.018  &kg/mol & \cite{Chang2010Chem}\\ 
$R$ & 8.314  & J/mol$\cdot$K & \cite{Chang2010Chem}\\ 
\hline  
\end{longtable}
\renewcommand{\arraystretch}{1}}

The parameters that should be estimated from experimental data include heat transfer coefficients, mass transfer coefficients and cake resistance, and desorption-related parameters (frequency factor and activation energy). Operating conditions that are not measured, e.g., wall temperatures, could also be estimated from data.

\section{Numerical Methods} \label{ch2-sec:Numeric}
This section comprehensively describes the numerical techniques required for solving the model equations efficiently. For the freezing model, since the lumped capacity approximation is used, the resulting equations are ordinary differential equations (ODEs). In \cite{Deck2022FreezingLumped,Deck2024Freezing2D}, the ODEs were solved numerically using an explicit scheme with fixed time steps. The probability of nucleation (Equation \eqref{ch2-eq:prob_nuc}) was calculated at every time step in a discrete fashion. From the computational perspective, this strategy is not the most efficient approach as it requires small step size to ensure stability. In this work, we rely on an implicit method with adaptive time steps; MATLAB's \texttt{ode15s} is selected. Nevertheless, since Equation \eqref{ch2-eq:prob_nuc} is an algebraic equation that depends on the time step $\Delta t$, solving an ODE system coupled with Equation \eqref{ch2-eq:prob_nuc} requires fixing $\Delta t$ via a predefined time span and then running the solver repeatedly and sequentially for each time step, with the probability of nucleation calculated at the end of each run, until the first nucleation is detected. A more efficient strategy is to consider a continuous version of Equation \eqref{ch2-eq:prob_nuc}, which is
\begin{equation} \label{ch2-eq:prob_con}
    P = 1 - \exp \left(\int_{t_0} ^ t -\lambda dt\right),
\end{equation}
where $t_0$ is the initial time and $P(t=t_0) = 0$. Differentiate Equation \eqref{ch2-eq:prob_con} results in
\begin{equation} \label{ch2-eq:prob_con2}
    \frac{dP}{dt}= \lambda(1-P) .
\end{equation}
The resulting ODE, Equation \eqref{ch2-eq:prob_con2}, can now be coupled with a system of ODEs describing heat and mass transfer in the process (e.g., during preconditioning and VISF), which can then be integrated using the selected ODE solver. In MATLAB, the Event function can be used to detect the time of first nucleation, i.e., when the probability exceeds the given value. With this strategy, our freezing model can be integrated continuously in a single run (i.e., without running the solver repeatedly). Consequently, the model equations can be solved within 0.1 s on a normal laptop, which is highly efficient.

For the PDEs in the primary drying and secondary drying models, the method of lines \cite{Schiesser1991MOL} is recommended. The method of lines consists of two steps. First, the PDEs are discretized spatially to produce a system of ODEs. Second, the resulting ODEs are integrated with a proper ODE solver (MATLAB's \texttt{ode15s} in this work). For the primary drying model, first consider Equation \eqref{ch2-eq:1st_energy}. As the equation involves a moving interface $S(t)$, define a new variable
\begin{equation}
    \xi = \frac{z-S}{H-S},
\end{equation}
where $\xi$ is the dimensionless position with respect to the moving interface (aka sublimation front). This dimensionless position varies from 0 to 1, in which $\xi=0$ at $z=S$ and $\xi=1$ at $z=H$.  Consequently, Equations \eqref{ch2-eq:1st_energy}, \eqref{ch2-eq:1st_bcbottom} and \eqref{ch2-eq:1st_bctop} become
\begin{gather}
    \rho_\textrm{f}\, C_{p,\textrm{f}}\,\dfrac{\partial T}{\partial t} = \frac{k_\textrm{f}}{(H-S)^2}\dfrac{\partial^2 T}{\partial \xi^2} - \frac{\rho_\textrm{f}\, C_{p,\textrm{f}}\,(\xi-1)}{H-S}\frac{dS}{dt}\frac{\partial T}{\partial \xi} + \frac{Q_\textrm{rad}}{V_\textrm{f}}, \qquad  0 < \xi < 1, \label{ch2-eq:1st_discretized1} \\
    -\frac{k_\textrm{f}}{H-S}\dfrac{\partial T}{\partial \xi} = h_\textrm{b}(T-T_\textrm{b}),  \qquad \xi=1, \label{ch2-eq:1st_discretized2} \\
    N_\textrm{w}\Delta H_\textrm{sub}  =  \frac{k_\textrm{f}}{H-S}\dfrac{\partial T}{\partial \xi} + \sigma\mathcal{F}_\textrm{s1}(T_\textrm{u}^4-T^4), \qquad \xi=0. \label{ch2-eq:1st_discretized3} 
\end{gather}
To spatially discretize the PDE, define
\begin{gather}
    \Delta \xi = \frac{1}{n_z-1}, \\
    \xi = (j-1) \Delta \xi,
\end{gather}
where $n_z$ is the number of grid points, $\Delta \xi$ is the distance between each grid point, and $j$ is the integer index for discretization, in which $j=1$ at $z=S$ and $j=n_z$ at $z=H$. Discretizing Equations \eqref{ch2-eq:1st_discretized1}--\eqref{ch2-eq:1st_discretized3} using the second-order finite difference scheme results in
\begin{gather}
    \begin{split}
         \dfrac{dT_j}{dt} = \frac{k_\textrm{f}}{\rho_\textrm{f} C_{p,\textrm{f}}(H-S)^2}\!\left(\dfrac{T_{j+1}-2T_j+T_{j-1}}{\Delta \xi^2}\right)\! - \frac{j \Delta \xi-1}{H-S}\frac{dS}{dt}\!\left(\dfrac{T_{j+1}-T_{j-1}}{2\Delta\xi}\right)\! + \frac{Q_\textrm{rad}}{\rho_\textrm{f} C_{p,\textrm{f}}V_\textrm{f}}, \\ \qquad  j=1,2,\dots,n_z, \label{ch2-eq:1st_discretized4}   
    \end{split} \\
    -\frac{k_\textrm{f}}{H-S}\!\left(\dfrac{T_{j+1}-T_{j-1}}{2\Delta\xi}\right)\! = h_\textrm{b}(T_{j}-T_\textrm{b}),  \qquad j=n_z, \label{ch2-eq:1st_discretized5} \\
    N_\textrm{w}\Delta H_\textrm{sub}  =  \frac{k_\textrm{f}}{H-S}\!\left(\dfrac{T_{j+1}-T_{j-1}}{2\Delta\xi}\right)\! + \sigma\mathcal{F}_\textrm{s1}\! \left(T_\textrm{u}^4-T_j^4\right), \qquad j=1, \label{ch2-eq:1st_discretized6} 
\end{gather}
where $T_j$ is the product temperature at position $j$. The ghost point at $j=n_z+1$ in Equation \eqref{ch2-eq:1st_discretized5} can be eliminated by substituting $T_{j=n_z+1}$ into Equation \eqref{ch2-eq:1st_discretized4} for $j=n_z$. Similarly, $T_{j=0}$ in Equation \eqref{ch2-eq:1st_discretized6}  can be eliminated by substituting $T_{j=0}$ into Equation \eqref{ch2-eq:1st_discretized4} for $j=1$. The above nondimensionalization and discretization techniques result in a system of nonlinear ODEs defined on a moving-grid system.

Numerical treatment for the secondary drying model is simpler than those described for the primary drying model as there is no moving interface. Define
\begin{gather}
    \Delta z = \frac{H}{n_z-1}, \\
    z = (j-1) \Delta z,
\end{gather}
where $\Delta z$ is the distance between each grid point, $j=1$ at $z=0$, and $j=n_z$ at $z=H$. Discretizing Equations \eqref{ch2-eq:2nd_energy}, \eqref{ch2-eq:2nd_desorption}, \eqref{ch2-eq:2nd_bctbottom}, and \eqref{ch2-eq:2nd_bctop} using the second-order finite difference scheme results in
\begin{gather}
    \frac{dT_j}{dt}  = \frac{k_\textrm{e}}{\rho_\textrm{e} C_{p,\textrm{e}}}\!\left(\frac{T_{j+1}-2T_j+T_{j-1}}{\Delta z^2}\right)\! + \frac{\rho_d\Delta H_\textrm{des}}{\rho_\textrm{e} C_{p,\textrm{e}}}\!\left(\dfrac{dc_{\textrm{w},j}} {dt}\right)\! + \frac{Q_\textrm{rad}}{\rho_\textrm{e} C_{p,\textrm{e}}V_\textrm{e}}, \qquad j = 1,2,\dots,n_z, \label{ch2-eq:2nd_discretized1} \\
    \dfrac{dc_{\textrm{w},j}}{dt} = -f_\textrm{a}e^{-E_\textrm{a}/RT_j} c_{\textrm{w},j}, \qquad  j = 1,2,\dots,n_z, \label{ch2-eq:2nd_discretized2}\\
    -k_\textrm{e}\!\left(\dfrac{T_{j+1}-T_{j-1}}{2\Delta z}\right) \!= h_\textrm{b}(T_{j}-T_\textrm{b}),  \qquad j=n_z, \label{ch2-eq:2nd_discretized3} \\
    -k_\textrm{e}\!\left(\dfrac{T_{j+1}-T_{j-1}}{2\Delta z}\right) \!= \sigma\mathcal{F}_\textrm{s1} \!\left(T_\textrm{u}^4-T_j^4\right),  \qquad j=1, \label{ch2-eq:2nd_discretized4}
\end{gather}
where the ghost points can be treated similarly as done for the primary drying model. 

All simulations, calculations, and results presented in this article were performed and generated using MATLAB R2023a. With the aforementioned numerical methods, our model can be simulated accurately within 1 s on a normal laptop. The model, data, and MATLAB code used in this work are made available (see the Data Availability section). 

\section{Results and Discussion} \label{ch2-sec:Results}

\subsection{Model validation} \label{ch2-sec:ModelValidation}
The mechanistic model presented in Section \ref{ch2-sec:Model} is validated using the experimental data reported in \cite{Capozzi2019ContLyo_SuspendedVials}. Our model validation is carried out for each lyophilization step separately, with the specific parameters reported in Table \ref{ch2-tab:ParametersValidation}.

{\singlespacing \renewcommand{\arraystretch}{1.2}
\begin{longtable}[ht!]{llll} 
\caption{Parameters used for the model validation, including Cases 1, 2a, 2b, 3a, and 3b.} 
\label{ch2-tab:ParametersValidation} \\
\hline
\textbf{Symbol} &\textbf{Value} & \textbf{Unit} & \textbf{Source} \\
\hline
\multicolumn{4}{l}{\textbf{Freezing (no VISF)}}  \\ 
$T_0$ & 280 & K & \cite{Capozzi2019ContLyo_SuspendedVials} \\
$T_\textrm{c},T_\textrm{u}$ & 272  & K &  estimated from data in \cite{Capozzi2019ContLyo_SuspendedVials} \\
$T_\textrm{g}$ & see Figure \ref{fig:Validation}A  & K & \cite{Capozzi2019ContLyo_SuspendedVials} \\
$T_\textrm{n}$ & 263.18 & K & \cite{Capozzi2019ContLyo_SuspendedVials} \\
$h_\textrm{s1}$ & 7 & W/m$^2$$\cdot$K & estimated from data in \cite{Capozzi2019ContLyo_SuspendedVials} \\
$h_\textrm{s2}$ & 18 & W/m$^2$$\cdot$K & estimated from data in \cite{Capozzi2019ContLyo_SuspendedVials} \\
$h_\textrm{s3}$ & 15 & W/m$^2$$\cdot$K & estimated from data in \cite{Capozzi2019ContLyo_SuspendedVials} \\ \hline

\multicolumn{4}{l}{\textbf{Primary drying}}  \\ 
$T_0$ & 231 (Case 2a) & K & \cite{Capozzi2019ContLyo_SuspendedVials} \\
& 235 (Case 2b) & K & \cite{Capozzi2019ContLyo_SuspendedVials} \\
$T_\textrm{b} $ & 263 (Case 2a)  & K &  \cite{Capozzi2019ContLyo_SuspendedVials} \\
& 313 (Case 2b)  & K & \cite{Capozzi2019ContLyo_SuspendedVials} \\
$T_\textrm{c},T_\textrm{u}$ & 275  & K & estimated from data in \cite{Capozzi2019ContLyo_SuspendedVials} \\
$h_\textrm{b}$ & 16 & W/m$^2$$\cdot$K & estimated from data in \cite{Capozzi2019ContLyo_SuspendedVials} \\
$R_\textrm{p1}$ & 3.4$\times$10$^{7}$ & 1/s &  estimated from data in \cite{Capozzi2019ContLyo_SuspendedVials} \\
$R_\textrm{p2}$ & 1 & 1/m &  estimated from data in \cite{Capozzi2019ContLyo_SuspendedVials} \\ \hline

\multicolumn{4}{l}{\textbf{Secondary drying}}  \\ 
$c_\textrm{w,0}$ & 0.088 (Case 3a) & kg water/kg solid & \cite{Capozzi2019ContLyo_SuspendedVials} \\
& 0.075 (Case 3b) & kg water/kg solid & \cite{Capozzi2019ContLyo_SuspendedVials} \\
$T_0$ & 273 & K & \cite{Capozzi2019ContLyo_SuspendedVials} \\
$T_\textrm{b}$ & 293  & K &  \cite{Capozzi2019ContLyo_SuspendedVials} \\
$T_\textrm{c},T_\textrm{u}$ & 285 & K & estimated from data in \cite{Capozzi2019ContLyo_SuspendedVials}  \\
$f_\textrm{a}$  & 0.42 & 1/s & estimated from data in \cite{Capozzi2019ContLyo_SuspendedVials} \\ 
$E_\textrm{a}$ & 2.05$\times$10$^{4}$ & J/mol$\cdot$K & estimated from data in \cite{Capozzi2019ContLyo_SuspendedVials} \\
$h_\textrm{b}$ & 16 & W/m$^2$$\cdot$K & estimated from data in \cite{Capozzi2019ContLyo_SuspendedVials} \\
$V_\textrm{l}$ & 2$\times$10$^{-6}$ & m$^3$ & \cite{Capozzi2019ContLyo_SuspendedVials}  \\ \hline  
\end{longtable} 
\renewcommand{\arraystretch}{1}}

\begin{figure}[ht!]
\centering
    \includegraphics[scale=1]{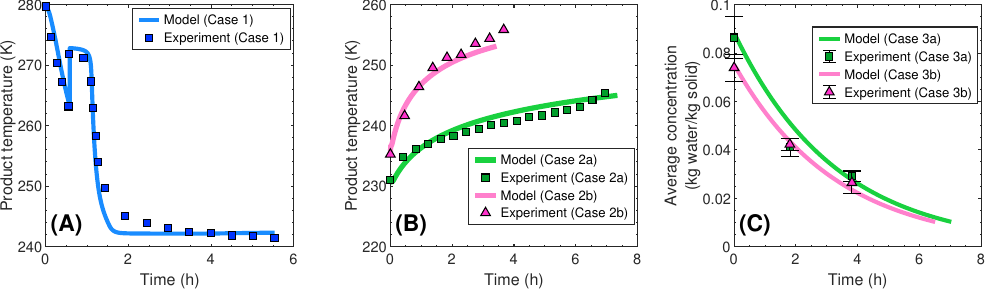}
    \caption{Model validation for the lyophilization of suspended vials using the experimental data from \cite{Capozzi2019ContLyo_SuspendedVials}. Panel A shows the model prediction and experimental data for the product temperature during the freezing step. Panel B shows the model prediction and experimental data for the product temperature (assumed to be measured at the bottom surface) during the primary drying step. The maximum shelf temperatures for Cases 2a and 2b are 263 and 313 K, respectively. Panel C shows the model prediction and experimental data for the average concentration of bound water during the secondary drying step. The initial concentrations for Cases 3a and 3b are 0.088 and 0.075 kg water/kg solid, respectively.} 
    \label{fig:Validation}      
\end{figure}

For the freezing step, the simulated temperature and freezing time agree well with the experimental data (Figure \ref{fig:Validation}A). The product temperature decreases from its initial value of 280 K to the nucleation temperature of about 263 K, where nucleation starts. Upon the first nucleation, the temperature instantaneously rises up from 263 K to the freezing point of about 272 K. The temperature slowly decreases during the solidification phase for about 30 min, following the freezing-point depression. Subsequently, the temperature decreases faster and reaches equilibrium with the environment (chamber wall and cryogenic gas), with the simulated temperature approaching the equilibrium slightly faster than the actual temperature.

The model accurately predicts the evolution of the product temperature and drying time during primary drying for both low (263 K) and high (313 K) shelf temperatures (Figure \ref{fig:Validation}B). The temperature rises up rapidly at the beginning and gradually increases at a slower rate toward the end of the process. This is because the driving force for sublimation is initially low, and so most of the heat input is used to increase the temperature. As time progresses, the driving force for sublimation is larger; thus, most of the heat input is used for sublimation. The maximum error in temperature prediction is about 2--3 K, while the error in drying time prediction is about 0.2 h for Case 2a and 0.4 h for Case 2b ($\sim$6\%). This error could be partly because of the uncertainty in cake resistance, which results from the stochastic nature of the freezing process. The datasets in Figure \ref{fig:Validation}B, which represent the lower and upper limits of the operating boundaries, demonstrate the validity and generalizability of the model across the full range of operating conditions.

Similar to primary drying, in secondary drying, we consider two different concentration levels for the bound water (residual moisture): 0.088 and 0.075 kg water/kg solid. The simulated concentration of bound water closely aligns with the experimental data for both concentration levels (Figure \ref{fig:Validation}C). The concentration decreases exponentially from its initial value to the final concentration of 0.01 kg water/kg solid, the threshold set for terminating the simulations. This concentration profile is expected given the linear driving force model is used to describe the desorption process. It is important to note that the data used in Figure \ref{fig:Validation}C are based on the shelf temperature of 293 K. To rigorously estimate the values of $f_\textrm{a}$ and $E_\textrm{a}$, it is recommended to obtain the concentration profiles at various shelf temperatures so that the effects of temperature on the desorption process can be quantified.  

Overall, our model is able to accurately predict the time evolution of the critical process parameters for all three steps of lyophilization. The maximum deviation between the predicted and measured temperature is about 3 K, which is about 1\% compared to the absolute operating temperature used in lyophilization. This error is smaller than the measurement noise of some non-invasive temperature sensors, e.g., thermal imaging cameras \cite{Srisuma2023Analytical} For the concentration of bound water, the maximum error is less than 0.01 kg water/kg solid, which is typically the threshold for identifying the endpoint of secondary drying \cite{Fissore2018Review} Therefore, the accuracy of our model is sufficient for general process design, optimization, and control.

We note that, in the suspended-vial lyophilization considered in this article, every single vial moves through the process following the same trajectory as shown in Figure \ref{fig:Vials}B, ensuring identical heat transfer conditions as a function of time for all vials. For batch lyophilization designs where the vials are distributed in different locations (Figure \ref{fig:Vials}A), the evolution of the critical process parameters in each vial also varies. For instance, the outermost vials in a batch lyophilizer dry the fastest due to thermal radiation from the environment, while the thermal radiation effect is much weaker on the inner vials \cite{Srisuma2024Rad}.

\subsection{Simulation of a complete lyophilization cycle} \label{ch2-sec:CompleteLyo}
One of the important aspects of continuous manufacturing is to ensure that the process is operated smoothly, optimally, and safely without human intervention, to minimize downtime and maximize production. In the context of continuous lyophilization, our model can be used to study the evolution of critical process parameters -- e.g., temperature, moisture content -- throughout the process at various operating conditions, which can help guide process design and optimization. 

This work not only presents the first mechanistic model for continuous lyophilization of suspended vials, but also is one of the very few studies that develops a complete model incorporating all three steps of lyophilization, including freezing, primary drying, and secondary drying. This is important specifically for continuous manufacturing where the entire process should be designed and optimized simultaneously.

\begin{figure}[ht!]
\centering
    \includegraphics[scale=1]{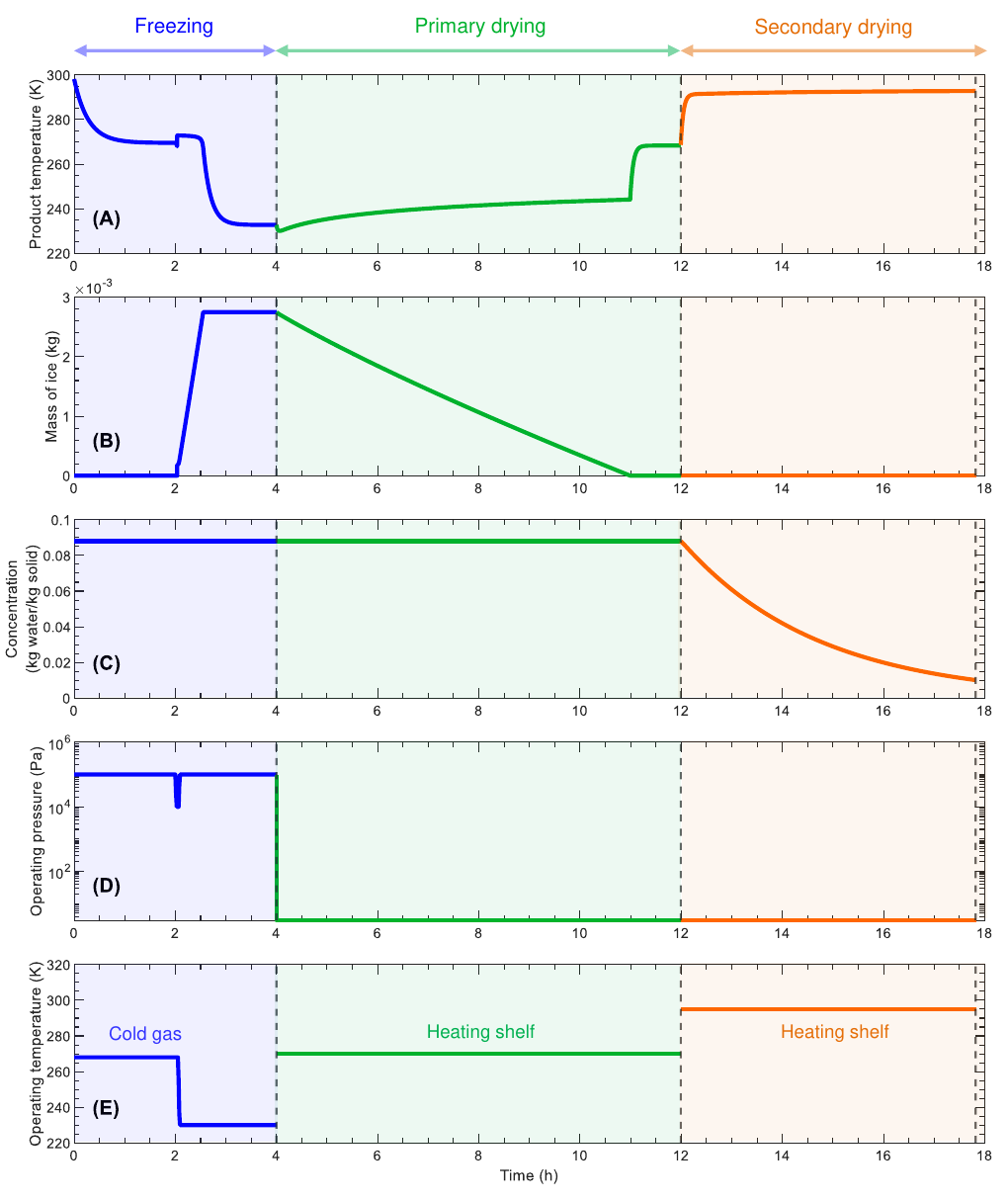}
    \caption{Complete simulation for the suspended-vial continuous lyophilization process showing the time evolution of the (A) product average temperature, (B) total mass of ice, (C) average concentration of bound water, (D) operating pressure, and (E) operating temperature.} 
    \label{fig:Complete_lyo}
\end{figure}

With the default model parameters in Table \ref{ch2-tab:DefaultParameters}, a complete continuous lyophilization cycle can be simulated, including (1) freezing with controlled nucleation via VISF, (2) primary drying, and (3) secondary drying (Figure \ref{fig:Complete_lyo}). The time evolution of the product temperature (Figure \ref{fig:Complete_lyo}A), mass of ice (Figure \ref{fig:Complete_lyo}B), and concentration of bound water (Figure \ref{fig:Complete_lyo}C) can be obtained, given the operating pressure (Figure \ref{fig:Complete_lyo}D) and temperature (Figure \ref{fig:Complete_lyo}E). This information could be useful for various purposes. For example, it can guide process and equipment design concerning the chamber sizing and velocity profile of each vial, ensuring that the target residence time is met for each chamber. Simulation results can also be used to help identify pressure and temperature profiles to achieve some specific objectives, e.g., minimization of drying time. It is important to note that the entire simulation shown in Figure \ref{fig:Complete_lyo} can be computed in less than 1 s on a normal laptop, which makes the model very practical to be implemented for any purposes.

\subsection{Visualizing spatiotemporal data} \label{ch2-sec:Spatial}
As discussed in Section \ref{ch2-sec:Strategies}, one of the key considerations during the drying steps is to ensure that the product temperature at any location does not exceed the upper limit. Consequently, spatiotemporal data are needed. Spatial variation of temperature and concentration in the product are dependent on several factors. For instance, temperature gradients are larger when the sample thickness and heat transfer coefficient increase (see the analysis in Appendix \ref{app:A}). 

\begin{figure}[ht!]
\centering
    \includegraphics[scale=1]{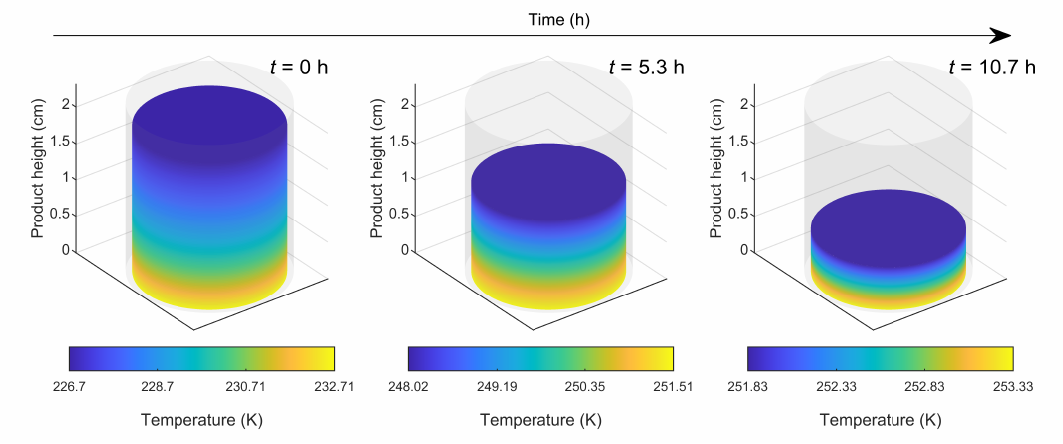}
    \caption{Spatiotemporal evolution of the product temperature and sublimation front during primary drying. Only the frozen region is shown in the figure.} 
    \label{fig:Lyo3D_1}      
\end{figure}

\begin{figure}[ht!]
\centering
    \includegraphics[scale=1]{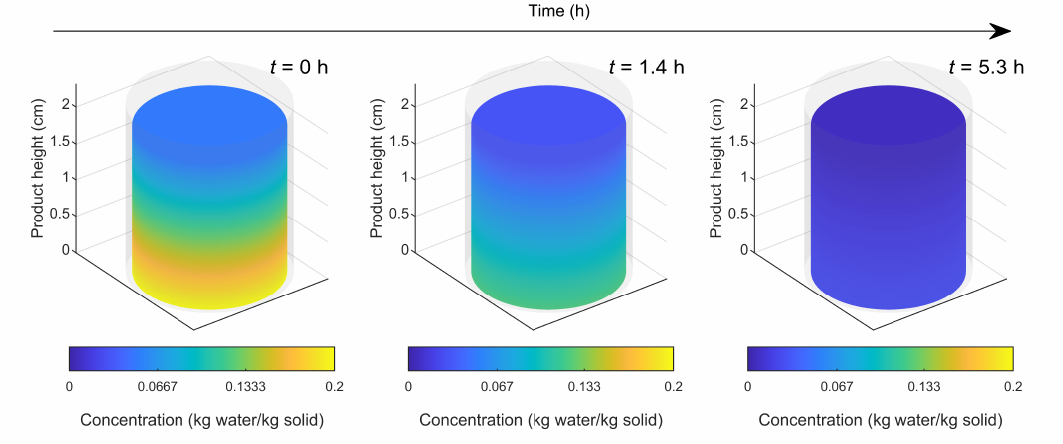}
    \caption{Spatiotemporal evolution of the concentration of bound water during secondary drying. The initial concentration is assumed to vary linearly from 5\% (0.05 kg water/solid) at the top to 20\% (0.2 kg water/kg solid) at the bottom. } 
    \label{fig:Lyo3D_2}      
\end{figure}

To demonstrate, we set the heat transfer coefficient and sample thickness to 30 W/m$^2$$\cdot$K and 0.02 m \cite{Sheehan1998Modeling}, respectively. These values are noticeably higher than the default values in Table \ref{ch2-tab:DefaultParameters}, and so the resulting gradients are more significant. Examples of spatiotemporal data obtained from our model are shown in Figures \ref{fig:Lyo3D_1} and \ref{fig:Lyo3D_2}. During primary drying, the temperature gradient is about 6 K at the beginning and gradually decreases as the sublimation front recedes (Figure \ref{fig:Lyo3D_1}). A similar behavior can also be observed for the concentration of bound water in secondary drying, except that there is no sublimation front (aka moving interface).

\subsection{Understanding and optimizing VISF} \label{ch2-sec:VISFresult}
A key benefit of mechanistic models is that they can sometimes provide important insights about the process when the experimental data are limited. This section employs the model and simulation to study the freezing step, which usually receives little attention compared to the drying steps. Spontaneous (aka uncontrolled) ice nucleation was already investigated in detail by \cite{Deck2022FreezingLumped,Deck2024Freezing2D}. However, uncontrolled nucleation is not ideal for continuous lyophilization as described in \cite{Capozzi2019ContLyo_SuspendedVials}. Therefore, we focus on controlled nucleation, with the VISF technique as used in \cite{Capozzi2019ContLyo_SuspendedVials}. The input and parameter data are given in Tables \ref{ch2-tab:VISF} and \ref{ch2-tab:VISF_exp}.

\begin{table}[ht!]
\renewcommand{\arraystretch}{1.2}
\caption{Parameters used for the modeling of VISF in Figures \ref{fig:VISF}A--D.} 
\label{ch2-tab:VISF}
\centering
\begin{threeparttable}
 \renewcommand{\arraystretch}{1.2}
\begin{tabular}{p{7em}  p{6em}  p{7em}}
\hline
\textbf{Symbol} &\textbf{Value} & \textbf{Unit} \\
\hline
$p_\textrm{t}$ & see\tnote{1} & Pa \\
$T_0$ & 280 & K \\
$T_\textrm{g},T_\textrm{c},T_\textrm{u}$ & see\tnote{2} & K \\
$h_\textrm{s2}$ & 60 &  W/m$^2$$\cdot$K \\
$h_\textrm{s3}$ & 60 &  W/m$^2$$\cdot$K \\
\hline  
\end{tabular}
 \renewcommand{\arraystretch}{1}
\begin{tablenotes}
{\footnotesize 
\item[1]$p_\textrm{t}$ is initially constant at $10^5$ Pa and then decreases linearly to the VISF pressure (different values are used in this study) in 1 min after VISF starts at $t$ = 0.25 h.
\item[2]$T_\textrm{g},T_\textrm{c}$, and $T_\textrm{u}$ are initially constant at 268 K and then decrease linearly to 260 K in 1 min after VISF starts at $t$ = 0.25 h.}
\end{tablenotes}
\end{threeparttable}
\end{table}

\begin{table}[ht!]
\renewcommand{\arraystretch}{1.2}
\caption{Parameters used for the validation of our VISF model in Figure \ref{fig:VISF}E.} 
\label{ch2-tab:VISF_exp}
\centering
\begin{threeparttable}
 \renewcommand{\arraystretch}{1.2}
\begin{tabular}{llll}
\hline
\textbf{Symbol} &\textbf{Value} & \textbf{Unit} & \textbf{Source} \\
\hline
$p_\textrm{t}$ &  see Figure 3 in \cite{Capozzi2019ContLyo_SuspendedVials} & Pa & \cite{Capozzi2019ContLyo_SuspendedVials} \\
$T_0$ & 268.27 & K & \cite{Capozzi2019ContLyo_SuspendedVials} \\
$T_\textrm{g}$ & see Figure 3 in \cite{Capozzi2019ContLyo_SuspendedVials} & K & \cite{Capozzi2019ContLyo_SuspendedVials} \\
$T_\textrm{c},T_\textrm{u}$ & 282 & K & estimated from data in \cite{Capozzi2019ContLyo_SuspendedVials} \\
$h_\textrm{m}$ & 1.3$\times$10$^{-2}$ & kg/m$^2$$\cdot$s & estimated from data in \cite{Capozzi2019ContLyo_SuspendedVials} \\
\hline  
\end{tabular}
 \renewcommand{\arraystretch}{1}
\end{threeparttable}
\end{table}

\begin{figure}[ht!]
\centering
    \includegraphics[scale=1]{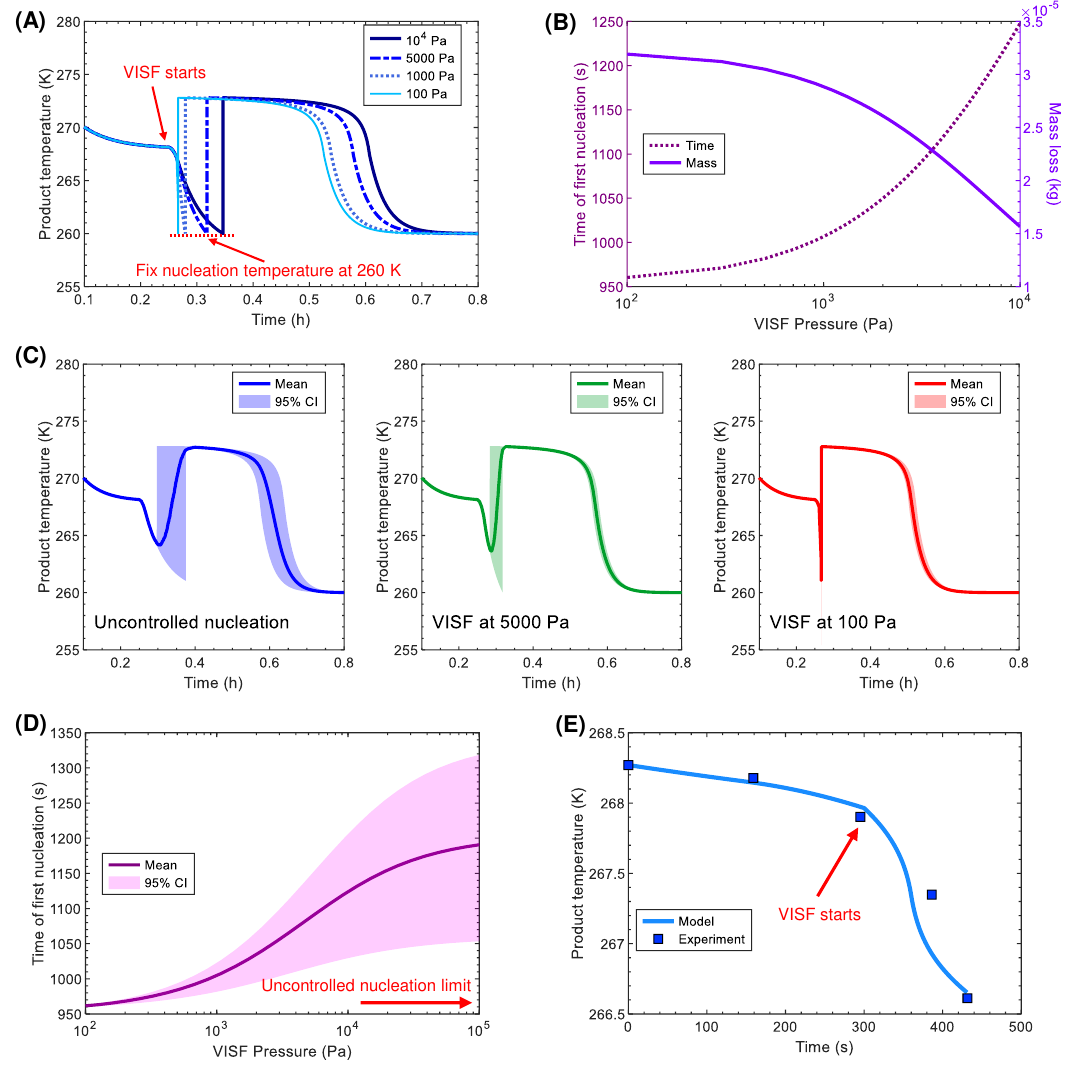}
    \caption{Controlled nucleation via vacuum induced surface freezing (VISF). (A) Product temperature profiles at different VISF pressures when the nucleation temperature is fixed at 260 K. (B) Nucleation time and mass loss due to evaporation at different VISF pressures when the nucleation temperature is fixed at 260 K. (C) Product temperature profiles for uncontrolled nucleation and VISF at 5000 Pa and 100 Pa considering stochastic ice nucleation. (D) Nucleation time at different VISF pressures considering stochastic ice nucleation. (E) Comparison between the simulated temperature profile during VISF and experimental data from \cite{Capozzi2019ContLyo_SuspendedVials}. Note that the nucleation time here is defined as the time when the first nucleation occurs.} 
    \label{fig:VISF}      
\end{figure}

Several experimental studies on VISF have been conduced in the literature. Here, we use our model and simulation results to provide insights into the process, and use that understanding to explain the previous experimental data. To clearly understand VISF, we first consider cases where that the nucleation temperature is fixed at 260 K; i.e., the effect of stochastic ice nucleation is excluded from the model (Figures \ref{fig:VISF}AB). In VISF, the key idea is to reduce the system pressure, typically from the atmospheric pressure ($10^5$ Pa) to the VISF pressure, to evaporate a small amount of liquid, resulting in a fast decrease in the product temperature and hence nucleation. The product temperature decreases faster with a lower VISF pressure (Figure \ref{fig:VISF}A). For example, when VISF starts at 0.25 h, the temperature drops from the initial value of 268 K to the nucleation temperature of 260 K almost instantaneously for the VISF pressure of 100 Pa, while it takes about 10 min when the VISF pressure is $10^4$ Pa. Consequently, the first nucleation occurs earlier at a lower VISF pressure (dash line in Figure \ref{fig:VISF}B). During VISF, a small amount of liquid evaporates, which increases with a decrease in the VISF pressure (solid line in Figure \ref{fig:VISF}B). This is expected because of a higher mass transfer driving force for evaporation at lower VISF pressure. In general, if VISF is carried out properly, the amount of liquid evaporating is small. In this case, the original liquid mass is about $2.9\times10^{-3}$ kg, and so the mass loss of about $3\times10^{-5}$ is almost negligible. 

Next, we include the effect of stochastic ice nucleation into our VISF model for a more realistic analysis (Figures \ref{fig:VISF}CD), with a Monte Carlo simulation of $10^4$ runs for proper statistics. Our simulation result show that VISF can significantly reduce the variation in the product temperature, compared to the uncontrolled nucleation case (Figure \ref{fig:VISF}C). A decrease in the VISF pressure also reduces the degree of variation. Besides the product temperature, The variation in the nucleation time\footnote{defined as the time when the first nucleation occurs} also decreases with the VISF pressure, with the first nucleation occurring earlier (Figure \ref{fig:VISF}D). These results indicate that the nucleation process is well controlled with VISF (i.e., less variation in the nucleation time and temperature), agreeing with general experimental observation in the literature. 

Finally, for model validation and implementation in real-world applications, the model can be fitted to experimental data obtained from the system of interest for more accurate results; the two important parameters are the mass transfer coefficient and heat transfer coefficient associated with evaporation. For example, by using the data from \cite{Capozzi2019ContLyo_SuspendedVials}, the temperature profile simulated by our model agrees well with the experimental data (Figure \ref{fig:VISF}E). 

The results presented in this section assume minimal vial-to-vial thermal interactions because the vials are arranged in a single row, in which all the vials experience the same heat transfer conditions throughout the process, as described in Section \ref{ch2-sec:heattransfer}. This condition is not true in conventional lyophilization where a large number of vials are arranged in a hexagonal or rectangular array. In such cases, vial-to-vial interactions could significantly affect the distribution of nucleation times and freezing rates. For example, when one vial nucleates, an increase in the temperature of that vial could slightly heat the adjacent vials, which subsequently delays the first nucleation of those vials. The effects of vial-to-vial interactions on freezing have been discussed in \cite{Deck2022FreezingLumped,Deck2024Freezing2D,Pisano2025Freezing}. In the context of modeling, these interactions can be incorporated straightforwardly by coupling our freezing model with the radiation network approach described in Section \ref{ch2-sec:heattransfer} and \cite{Srisuma2024Rad}. In any case, such effects become much less important in controlled nucleation.

In summary, this section demonstrates how the model and simulation results elucidate the role of VISF and its operating conditions in promoting controlled nucleation and influencing the nucleation process, improving the uniformity of the product. Furthermore, we show that the model can explain real data relatively well. Hence, our model can be used to understand and guide the design and optimization of the VISF method for controlling the nucleation process in lyophilization.

\subsection{Analysis of condenser failure}
Typical process modeling focuses on behaviors of the system during its normal operation where the process is operated steadily under the desired conditions. A better process model should be able to simulate important abnormal conditions, e.g., equipment failure, that could occur due to various unexpected scenarios, which is critical in continuous manufacturing as these abnormal operations could affect the reliability, availability, and maintainability (RAM) as well as the safety of the plant/process. 

In lyophilization, one of the crucial design considerations is associated with a condenser. During normal operation, the maximum capacity of a condenser must be higher than the rate of vapor production via sublimation, and choked flow should be avoided \cite{Fissore2018Review}. This operational constraint is to ensure that there will not be vapor accumulation in the equipment, which could subsequently lead to in a pressure increase. This scenario may not lead to safety-related issues because the operating pressure in lyophilization is low and the amount of ice/water is generally not high enough to create overpressure. Nevertheless, an increase in the total pressure reduces the driving force for sublimation (see Equation \eqref{ch2-eq:1st_flux}), which then prolongs the primary drying step. As a result, the final product quality could be affected. This section explores how to incorporate those dynamic behaviors into our model.

Most primary drying models, including our model, assume that the system pressure is well controlled, which is a reasonable assumption. To incorporate the condenser dynamics into the model, we consider cases where the total condenser capacity is not enough to condense the water vapor produced via sublimation during primary drying. In such cases, the mass balance for the water vapor in a chamber is \cite{Bano2020LumpedDrying} 
\begin{equation} \label{ch2-eq:pressure_balance}
    \frac{dp_\textrm{w,c}}{dt} = \frac{(j_\textrm{w}-j_\textrm{w,max})R\overline{T}}{V_\textrm{c}M_\textrm{w}},
\end{equation}
where $j_\textrm{w}$ is the total mass flow rate of water vapor resulting from sublimation, $j_\textrm{w,max}$ is the maximum condenser capacity, $V_\textrm{c}$ is the chamber volume, and $T_\textrm{avg}$ is the average temperature in the chamber assumed to be constant for simplification. The mass flow rate of water vapor can be calculated by 
\begin{equation} \label{ch2-eq:massflow_vapor}
    j_\textrm{w} = n_\textrm{vial}A_zN_w,
\end{equation}
where $n_\textrm{vial}$ is the number of vials in the chamber. By coupling Equations \eqref{ch2-eq:pressure_balance} and \eqref{ch2-eq:massflow_vapor} with the primary drying model developed in Section \ref{ch2-sec:Model1stDrying}, the effects of condenser failure or choked flow can be quantified. 

\begin{figure}[ht!]
\centering
    \includegraphics[scale=1]{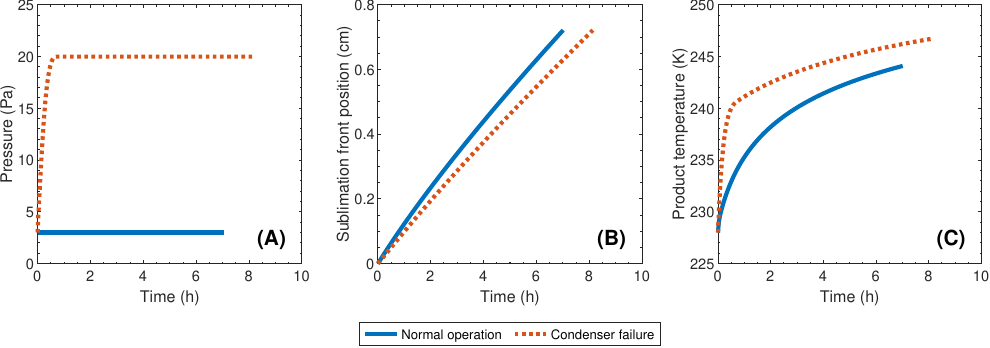}
    \caption{Simulation results comparing the normal operation and condenser failure scenario for the (A) system pressure, (B) sublimation front position, and (C) product average temperature during primary drying.} 
    \label{fig:Choked}      
\end{figure}

With the specific parameters given in Table \ref{ch2-tab:ParametersChoked}, the total pressure increases from the normal operating value of 3 Pa to 20 Pa (Figure \ref{fig:Choked}A). This pressure increase results from the total vapor flow exceeding the maximum condenser capacity, thus vapor accumulation in the chamber. When the system pressure increases, the driving force for sublimation (also the sublimation flux) decreases. After about 1 h, the system pressure becomes constant, indicating the equalization of the sublimation flux and condenser capacity. In this abnormal operation, the drying time and product temperature are higher than for the normal operation (Figures \ref{fig:Choked}BC). 

\begin{table}[ht!]
\renewcommand{\arraystretch}{1.2}
\centering
\caption{Parameters used for the condenser failure analysis.} 
\label{ch2-tab:ParametersChoked}
\begin{tabular}[ht!]{llll} 
\hline
\textbf{Symbol} &\textbf{Value} & \textbf{Unit} & \textbf{Source} \\
\hline
$\overline{T}$ & 260 & K & assumed to be constant \\
$n_\textrm{vial}$ & 200 & -- & -- \\
$V_\textrm{c}$ & 0.118 & m$^3$ & \cite{Bano2020LumpedDrying} \\
$j_\textrm{w,max}$ & 1.8$\times$10$^{-5}$ & m$^3$/s & \cite{Bano2020LumpedDrying} \\
\hline  
\end{tabular}
\renewcommand{\arraystretch}{1}
\end{table}

\section{Conclusion} \label{ch2-sec:Conclusion}
This article proposes the first mechanistic model for continuous lyophilization, in which the state-of-the-art suspended-vial technology is considered. The model can simulate the entire lyophilization process, including the freezing, primary drying, and secondary drying steps. The freezing model considers preconditioning, vacuum-induced surface freezing (VISF), spontaneous and controlled nucleation, solidification, and cooling. The primary drying model captures heat transfer in the frozen region and mass transfer resulting from sublimation of ice crystals at the sublimation front. The secondary drying model describes simultaneous heat transfer in the dried region and desorption of bound water. The overall model can describe the evolution of product temperature, ice/water fraction, and residual moisture throughout the process.

The proposed model is validated for all three steps including freezing, primary drying, and secondary drying, in which the model predictions are consistent with the experimental measurements in all cases. The validated model is demonstrated for a variety of applications related to process design and optimization. Every single simulation can be run within less than 1 s on a normal laptop, allowing the model to be employed for any purpose. The framework and results presented in this work are suitable for guiding the design and development of future continuous lyophilization technology. 

With a well-developed mechanistic model, future work could consider utilizing the model for several important tasks, e.g., state estimation, optimal control design, and uncertainty quantification, which will ultimately support the development of a high-quality digital twin for continuous lyophilization to advance the manufacturing process.

\section*{Data and Code Availability}  \label{ch2-sec:code}
Software and data used in this work are available at \url{https://github.com/PrakitrSrisuma/ContLyo-modeling}. All simulations, calculations, and results presented in this article were performed and generated using MATLAB, but an equivalent Julia implementation of the code is also provided.

\section*{Acknowledgments} 
RDB and PS were supported by the U.S. Food and Drug Administration under the FDA BAA-22-00123 program, Award Number 75F40122C00200. The authors would like to thank Pitipat Wongsittikan for helping optimize part of the Julia code.

\medskip

%


\section*{Appendices}
\appendix
\renewcommand\thefigure{\thesection.\arabic{figure}} 
\renewcommand\theequation{\thesection.\arabic{equation}}

\setcounter{figure}{0} 
\setcounter{equation}{0} 

\section{Analysis of Transport Phenomena} \label{app:A}

\subsection{Biot Number and Spatial Dimensions}
As discussed in the main text, the freezing step is modeled using the lumped capacity method, whereas the drying steps are analyzed in one spatial dimension, a vertical direction. Various studies have shown via simulation that variation in the radial direction is negligible \cite{Sheehan1998Modeling,Deck2024Freezing2D}. This section provides a detailed analysis from the theoretical perspective to justify this assumption and briefly discusses strategies to modify the model for situations where this assumption is not true.

Introduce the Biot number
\begin{equation}
    \textrm{Bi} = \frac{hL}{k},
\end{equation}
where $L$ is the characteristic length or length scale, $h$ is the heat transfer coefficient, and $k$ is the thermal conductivity. In general, the length scale is defined as the ratio of the volume to the surface area, which results in $r_\textrm{o}/2$ for a long cylinder. However, for conservative analysis, it is recommended to use the actual radius $r_\textrm{o}$ instead \cite{Incropera2007HeatTransfer}. The Biot number is defined as the ratio of the internal conduction resistance to the external convection resistance, and so a small Bi value implies that the resulting temperature gradient should also be small \cite{Mills1995HeatTransfer}. In various heat transfer textbooks, the value of 0.1 is recommended as a guideline; if Bi $< 0.1$, a lumped capacity model can be used. However, in practice, a lumped capacity model could still be valid even for Bi $> 0.1$, depending on the system, which is analyzed here.

\begin{figure}[ht!]
\centering
\includegraphics[scale=1]{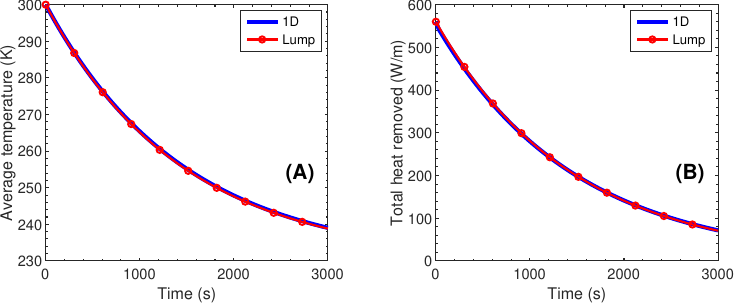}
 \caption{Comparison between the 1D and lumped capacity models for the (A) volume-averaged temperature and (B) total heat removed for the default model parameters.} 
\label{fig:Biot_base}
\end{figure}

This work considers freeze drying of unit doses, with a vial modeled as a cylinder. Since our focus in on the radial heat transfer, we consider an infinite cylinder to minimize any contribution from the vertical heat transfer. Analytical solutions to the 1D heat transfer problem in an infinite cylinder can be found in \cite{Mills1995HeatTransfer}; we denote this as the 1D model. The properties of ice are used in this analysis because the product is in a frozen state for most of the time. First, consider the base case in which the default parameter values in Table \ref{ch2-tab:DefaultParameters} are used. In this case, $h_\textrm{s3} = 8$ W/m$^2$$\cdot$K, $k_\textrm{i} = 2.25$ W/m$\cdot$K and $L = 0.012$ m, resulting Bi = 0.043. The initial temperature is set to 300 K, with the cold gas temperature set to 230 K, a similar condition used in the freezing step. Since the Biot number is less than 0.1, the lumped capacity model should be give accurate results. As expected, the temperature and total heat removed predicted from both models are almost identical (Figure \ref{fig:Biot_base}).

\begin{figure}[ht!]
\centering
\includegraphics[scale=1]{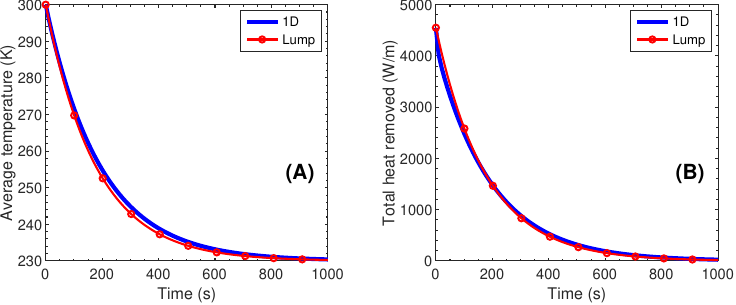}
\caption{Comparison between the 1D and lumped capacity model for the (A) volume-averaged temperature and (B) total heat removed for the conservative case.} 
\label{fig:Biot_max}      
\end{figure}

Next consider a more conservative case. The Biot number increases with an increase in the heat transfer coefficient. Heat transfer coefficients in freeze drying reported in the literature typically range between 4 and 30 W/m$^2$$\cdot$K \cite{Sheehan1998Modeling,Pikal2005Model,Veraldi2008SimBatchModels}. The highest value, to our knowledge, is 65 W/m$^2$$\cdot$K \cite{Hottot2006Freezing}. With the heat transfer coefficient of 65 W/m$^2$$\cdot$K, Bi = 0.35, which is now higher than 0.1. However, the results predicted by the 1D and lumped capacity models are still very similar (Figure \ref{fig:Biot_max}), although the difference is more noticeable than that observed in the base case.

\begin{figure}[ht!]
\centering
\includegraphics[scale=1]{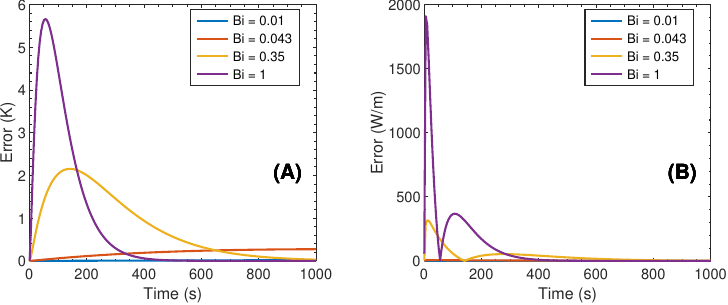}
\caption {Errors between the 1D and lumped capacity model for the (A) volume-averaged temperature and (B) total heat removed for different Bi.} 
\label{fig:Biot_all}      
\end{figure}

Finally, the errors (differences) between both models for different values of Bi are shown in Figure \ref{fig:Biot_all}. Of course, if the Biot is too high, the lumped capacity model will not be accurate. That situation could happen, e.g., when a very large vial is used or the heat transfer coefficient is much higher than usual. In such cases, a high-fidelity model might be needed \cite{Deck2024Freezing2D}. However, the analysis in this section suggests that a lumped capacity model is highly accurate for typical lyophilization applications, in particular for unit doses. 

\subsection{Mass Transport in the Dried Product}
In secondary drying, the two important mechanisms for mass transport are (1) vapor phase or water vapor transport through the porous dried product and (2) desorption of bound water from the solid surface of the dried product. Previous simulation and experimental data suggest that desorption is the rate-limiting step \cite{Pikal1990SecDrying_Kinetics,Fissore2011SecDryingMonitor,Fissore2015Review,Sahni2017Simplified,Yoon2021Sec0D1D3D,Srisuma2024Obs}, and so the equations/terms related to vapor phase transport are typically omitted from recently published models, including our model. This section provides an analysis from the perspective of transport phenomena to justify this assumption.

First, consider the vapor phase transport through the porous dried product. Modeling the transport of liquids and gases in porous media is challenging as it could entail a variety of mechanisms besides ordinary diffusion (e.g., Knudsen diffusion), and thus no fully satisfactory theory currently exists \cite{Bird2002Transport}. Nevertheless, relative success has been achieved by using the concept of an effective diffusivity; in this analysis, we follow the procedure described in \cite{Mills1995HeatTransfer}. Define the effective diffusivity $D_\textrm{e}$ by
\begin{equation} \label{appa-eq:EffDiff}
    \frac{1}{D_\textrm{e}} =  \frac{1}{D_\textrm{e,g}} +  \frac{1}{D_\textrm{e,K}},
\end{equation}
where $D_\textrm{e,g}$ is the effective diffusion coefficient for ordinary diffusion, i.e., diffusion of the molecules in the vapor/gas phase, $D_\textrm{e,K}$ is the effective diffusion coefficient for Knudsen diffusion, and $D_\textrm{e}$ is the effective diffusivity that combines the effects of ordinary diffusion and Knudsen diffusion. For ordinary diffusion,  
\begin{equation} \label{appa-eq:OrdDiff}
    D_\textrm{e,g} = \frac{\epsilon}{\tau} D_\textrm{g},
\end{equation}
where $D_\textrm{g}$ is the binary diffusion coefficient, e.g., water vapor in air, $\epsilon$ is the porosity, and $\tau$ is the tortuosity. For Knudsen diffusion, a similar expression can be used, that is,
\begin{equation} \label{appa-eq:KEffDiff}
    D_\textrm{e,K} = \frac{\epsilon}{\tau} D_\textrm{K},
\end{equation}
where $D_\textrm{K}$ is the Knudsen diffusion coefficient. This coefficient could be calculated from
\begin{equation} \label{appa-eq:KDiff}
    D_\textrm{K} = 97\bar{r}\sqrt{\frac{T}{M}},
\end{equation}
where $\bar{r}$ is the average pore radius, $T$ is the temperature, and $M$ is the molar mass in g/mol.

With the effective diffusivity calculated by Equations \eqref{appa-eq:EffDiff}--\eqref{appa-eq:KDiff}, the total mass flux can be calculated using Fick's law of diffusion, with the binary diffusion coefficient replaced by the effective diffusivity. Then, a continuity (mass balance) equation for the water vapor can be written and coupled to the mechanistic model proposed in the main paper. The key benefit of this technique is that the implementation closely resembles the approach for solving problems entailing ordinary diffusion of a binary mixture, which is relatively straightforward. Nevertheless, additional information is needed, including the porosity, tortuosity, and average pore radius. A more complicated strategy that is used in the literature is to apply the dusty gas model, which is derived from the Maxwell-Stefan Equation (see examples in \cite{Sadikoglu1997Modeling,Veraldi2008SimBatchModels}). This approach, however, requires many more parameters and equations.

Next, we compare the significance of vapor phase transport and desorption. Scale analysis can be used to analyze the contribution of each transport process (see various examples in \cite{Mills1995HeatTransfer}). Properly estimating the time scale of each phenomenon can help identify the key process that should be considered in the model and omit unimportant phenomena. The time scale (aka time constant) for mass diffusion can be approximated by \cite{Crank1979Diffusion,Mills1995HeatTransfer}
\begin{equation} \label{appa-eq:timediff}
    t_\text{dff} = \frac{L^2}{D_\textrm{e}},
\end{equation}
where $t_\text{dff}$ is the time scale for diffusion and $L$ is the proper length scale. Similarly, the time scale for desorption can be approximated by
\begin{equation} \label{appa-eq:timedes}
    t_\text{des} = \frac{1}{k_\textrm{d}},
\end{equation}
where $t_\text{des}$ is the time scale for desorption and $k_\textrm{d}$ is the desorption constant defined in Equation \eqref{ch2-eq:2nd_desorption}. By comparing $t_\text{dff}$ and $t_\text{des}$, the limiting step for mass transfer in the secondary drying step can be identified, without any simulation or experimental data. 

For numerical comparison, a set of parameter values are obtained from \cite{Liapis1994Original,Sheehan1998Modeling}, namely $\bar{r} = 5$$\times$$10^{-6}$ m, $\epsilon = 0.815$, $\tau = 1.2$, and $k_\textrm{d} = 7.8$$\times$$10^{-5}$ s$^{-1}$. The binary coefficient of water in air is obtained from \cite{Mills1995HeatTransfer}, which is $D_\textrm{g} = 1.97$$\times$$10^{-5}$ m$^2$/s at 256 K and 1 atm. In secondary drying, the operating pressure is much lower than 1 atm, and so the actual $D_\textrm{g}$ is lower as the diffusion coefficient of gas mixtures at low pressure is inversely proportional to the pressure \cite{Mills1995HeatTransfer,Bird2002Transport}, which makes our analysis more conservative. The length scale for this analysis is set to 0.01 m, a typical vial/sample size (thickness/diameter). Substituting these values into Equations \eqref{appa-eq:timediff} and \eqref{appa-eq:timedes} gives $t_\text{dff} = 7.5$ s and $t_\text{des} = 3.6 $ h. Obviously, the time scale for diffusion is much smaller than for desorption by several orders of magnitude, indicating that desorption is the rate-limiting step. A typical drying time of secondary drying is on the order of several hours, agreeing with the calculated $t_\text{des}$ of about 3.6 h. On the other hand, the time scale for diffusion is only 7.5 s, meaning that diffusion is much faster and so can be safely omitted from the model. Even some of the parameters are changed by an order of magnitude, this conclusion does not change.\footnote{This analysis holds for freeze drying of unit doses only; for freeze drying of bulk material (e.g., in food applications), the length scale of the product could be much larger, and so the conclusion might be different.}

Finally, there is another mass transfer mechanism that is not commonly discussed or modeled, namely diffusion of water in the solid matrix (not in the pores), i.e., solid-phase diffusion \cite{Pikal1990SecDrying_Kinetics}. Diffusion in the solid matrix is very slow, and so it could limit the mass transport in secondary drying. Modeling this phenomenon is challenging as it requires detailed knowledge of the solid matrix, particle size, and locations of bound water in the matrix, which is generally not available. A similar approximation to Equation \eqref{appa-eq:timediff} can be carried out. By using the reported length scale of about $5$$\times$$10^{-7}$ m and diffusion coefficient of water in the solid phase of about $7$$\times$$10^{-16}$ m$^2$/s \cite{Pikal1990SecDrying_Kinetics}, the time scale for solid-phase diffusion is about 10 min, which is still much lower than for desorption.

To summarize, this section details the scale analysis that can be efficiently used to evaluate the effects of various mass transfer mechanisms. For freeze drying of (bio)pharmaceuticals in unit doses, our analysis shows that desorption of bound water is the rate-limiting step for mass transfer, and so other mechanisms can be omitted from the model, agreeing with both simulation and experimental data in the literature.

\setcounter{figure}{0} 
\setcounter{equation}{0} 
\section{Supplemental Equations} \label{app:B}
This appendix summarizes equations that are not a crucial part of the model but required for calculating some model parameters. 

For the freezing model, if the total liquid volume $V_\textrm{l}$ and mass fraction of a solute $x_\textrm{s}$ are reported, the density of the liquid phase $\rho_\textrm{l}$ is
\begin{equation}\label{appb-eq:rho_precond}
    \rho_\textrm{l} = \frac{V_\textrm{l}}{\dfrac{x_\textrm{s}}{\rho_\textrm{s}} + \dfrac{1-x_\textrm{s}}{\rho_\textrm{w}}}.
\end{equation}
Consequently, $m_\textrm{s}$ and $m_\textrm{w,0}$ can be calculated by
\begin{gather}
    m_\textrm{s} = x_\textrm{s}\rho_\textrm{l}V_\textrm{l}, \\
    m_\textrm{w,0} = (1-x_\textrm{s})\rho_\textrm{l}V_\textrm{l}.    
\end{gather}
Then, Equation \eqref{ch2-eq:energy_precond} can be used as usual.

During the freezing step, the surface area $A_r$ varies with time due to changes in the total volume caused by phase transition. In general, the surface area can be calculated by
\begin{equation}
    A_r(t) = \frac{4\!\left(\dfrac{m_\textrm{s}}{\rho_\textrm{s}} + \dfrac{m_\textrm{w}(t)}{\rho_\textrm{w}} + \dfrac{m_\textrm{i}(t)}{\rho_\textrm{i}} \right)}{d},
\end{equation}
where the product is assumed to be a cylinder. 

For the primary drying model, the volume of the frozen region is
\begin{equation}
    V_\textrm{f} = \pi d^2(H-S). 
\end{equation}
The height/thickness of the product $H$ can be calculated by
\begin{equation}
    H = \frac{m_\textrm{s} + m_\textrm{w,0}}{\rho_\textrm{f}A_z}.
\end{equation}
The properties of the frozen region (subscript `f') can be calculated from those of the ice and solid material or solute from
\begin{gather}
    \rho_\textrm{f} = \frac{1}{\dfrac{x_\textrm{s}}{\rho_\textrm{s}} + \dfrac{1-x_\textrm{s}}{\rho_\textrm{i}}}, \\
    C_{p,\textrm{f}} = x_\textrm{s}C_{p,\textrm{s}} + (1-x_\textrm{s})C_{p,\textrm{i}}, \\
    k_\textrm{f} = x_\textrm{s}k_\textrm{s} + (1-x_\textrm{s})k_\textrm{i}.
\end{gather}
These expressions assume that the final mass of ice is equal to the initial mass of water (complete freezing), meaning that the amount of water loss due to evaporation during VISF and unfrozen water (bound water) is negligible, which is generally true. Note that some experimental primary drying studies report $\rho_\textrm{f},C_{p,\textrm{f}},k_\textrm{f},$ and/or $H$, in which case the above calculations may not be needed.

For the secondary drying model, the volume of the dried region is
\begin{equation}
    V_\textrm{e} = \pi d^2H,
\end{equation}
which is constant because there is no moving boundary (i.e., sublimation front) in this case.

\medskip
\renewcommand{\arraystretch}{1.2}
\begin{longtable}[ht!]{ l  l  } 
\caption*{} 
\label{Tab:ParametersSpecific}\\
\hline
\multicolumn{2}{l}{\textbf{Nomenclature}}   \\ 
\hline
\multicolumn{2}{l}{\textbf{Symbols}}   \\ 
$A$ & area (m$^2$) \\
$A_z$ & cross sectional area of the product (m$^2$) \\ 
$A_r$ & side surface area of the product (m$^2$) \\ 
$b_\textrm{n}$ & nucleation kinetics parameter (--) \\ 
$C_p$ & specific heat capacity (J/kg$\cdot$K) \\ 
$c_\textrm{w}$ & concentration of bound water (kg water/kg solid) \\ 
$c^*_\textrm{w}$ & equilibrium concentration of bound water (kg water/kg solid) \\ 
$D$ & diffusivity (m$^2$/s) \\ 
$D_\textrm{e}$ & effective diffusivity (m$^2$/s) \\ 
$D_\textrm{e,g}$ & effective diffusivity for ordinary diffusion of gases (m$^2$/s) \\ 
$D_\textrm{e,K}$ & effective diffusivity for Knudsen diffusion (m$^2$/s) \\ 
$D_\textrm{g}$ & binary diffusion coefficient of gases (m$^2$/s) \\ 
$D_\textrm{K}$ & Knudsen diffusion coefficient (m$^2$/s) \\ 
$d$ & vial diameter (m) \\ 
$E_\textrm{a}$ & activation energy (J/mol$\cdot$K) \\
$\mathcal{F}$ & transfer factor (--)\\ 
$F_{1-2}$ & view factor from surface 1 to surface 2 (--)\\ 
$f_\textrm{a}$ & frequency factor (1/s) \\
$H$ & product height/thickness (m) \\ 
$\Delta H_\textrm{des}$ & heat of desorption (J/kg) \\ 
$\Delta H_\textrm{fus}$ & heat of fusion (J/kg) \\ 
$\Delta H_\textrm{sub}$ & heat of sublimation (J/kg) \\ 
$\Delta H_\textrm{vap}$ & heat of vaporization (J/kg) \\ 
$h$ & heat transfer coefficient (W/m$^2$$\cdot$K) \\
$h_m$ & mass transfer coefficient (kg/m$^2$$\cdot$s)  \\ 
$j$ & integer index for discretization (--)  \\
$j_\textrm{w}$ & mass flow rate of water vapor (kg/s)  \\
$j_{\textrm{w,max}}$ & condenser maximum capacity (kg/s)  \\
$K_\textrm{f}$ & molal freezing-point depression constant (kg$\cdot$K/mol) \\ 
$k$ & thermal conductivity (W/m$\cdot$K) \\ 
$k_\textrm{d}$ & desorption constant (1/s) \\
$k_\textrm{n}$ & nucleation kinetics parameter (1/m$^3\cdot$s$\cdot$K$^{b_\textrm{n}})$ \\ 
$L$ & length scale (m) \\ 
$l$ & thickness of the ice layer (m) \\
$M$ & molar mass (kg/mol) \\ 
$m$ & mass (kg) \\ 
$N_\textrm{w}$ & mass flux of water (kg/m$^2\cdot$s)  \\ 
$n_\textrm{vial}$ & number of vials (--) \\ 
$n_z$ & number of grid points in the $z$-direction (--) \\ 
$P$ & probability \\
$p$ & partial pressure (Pa) \\
$p_\textrm{t}$ & total pressure (Pa) \\
$Q$ & heat transfer rate (W) \\ 
$Q_\textrm{rad}$ & radiative heat transfer rate (W) \\
$q$ & heat flux (W/m$^2$) \\ 
$R$ & gas constant (J/mol$\cdot$K) \\
$R_\textrm{p}$ & mass transfer resistance of the product during sublimation (m/s)  \\
$R_\textrm{p0}$ & parameter for the mass transfer resistance (m/s)  \\
$R_\textrm{p1}$ & parameter for the mass transfer resistance (1/s)  \\
$R_\textrm{p2}$ & parameter for the mass transfer resistance (1/m)  \\
$r$ & radial position (m)  \\
$r_\textrm{o}$ & radius of the vial (m)  \\
$\bar{r}$ & average pore radius (m)  \\
$S$ & sublimation front position (m) \\
$T$ & product temperature (K) \\
$\overline{T}$ & average chamber temperature (K) \\
$t$ & time (s) \\
$t_\textrm{0}$ & initial time (s) \\
$t_\textrm{f1}$ & time when preconditioning completes (s) \\
$t_\textrm{f2}$ & time when VISF completes (s) \\
$t_\textrm{f3}$ & time when first nucleation completes (s) \\
$t_\textrm{f4}$ & time when solidification completes (s) \\
$t_\textrm{f5}$ & time when final cooling completes (s) \\
$t_\textrm{d1}$ & time when primary drying completes (s) \\
$t_\textrm{d2}$ & time when secondary drying completes (s) \\
$t_\textrm{dff}$ & time scale for diffusion (s) \\
$t_\textrm{des}$ & time scale for desorption (s) \\
$V$ & volume (m$^3$) \\ 
$x$ & mass fraction (--) \\
$U$ & overall heat transfer coefficient (W/m$^2$$\cdot$K) \\
$z$ & position in vertical direction (m) \\
$\Delta z$ & distance between each grid point in the $z$-direction (m) \\
$\epsilon$ & porosity \\
$\varepsilon$ & emissivity (--) \\ 
$\xi$ & normalized vertical position (--) \\
$\Delta \xi$ & distance between each grid point in the $\xi$-direction (--) \\
$\rho$ & density (kg/m$^3$) \\ 
$\rho_\textrm{d}$ & density of the dried product (kg/m$^3$) \\ 
$\sigma$ & Stefan-Boltzmann constant (W/m$^2\cdot$K) \\ 
$\tau$ & tortuosity \\
  \hline
\multicolumn{2}{l}{\textbf{Common subscripts}}   \\ 
$0$ & initial condition \\
$1$ & surface 1 \\
$2$ & surface 2 \\
avg & average \\
c & chamber (environment), chamber wall (side)\\
b & bottom shelf \\
e & effective properties of the dried product (solid + gas) \\ 
f & frozen material, freezing \\ 
gl & glass vial \\
i & ice \\ 
l & liquid solution (water + solute/solid) \\ 
n & nucleation \\
s & solute/solid \\ 
sat & saturated condition \\
s1 & top surface of the product \\
s2 & bottom surface of the product \\
s3 & side surface of the product \\
u & upper surface of the chamber \\ 
w & water \\ 
in & inert gas \\ 
\hline  
\end{longtable}
\renewcommand{\arraystretch}{1}
\setcounter{table}{0}

\clearpage
\bibliographystyle{MSP}
\bibliography{reference}

@Incollection{Fissore2015Review,
    author={Fissore, Davide and Pisano, Roberto and Barresi, Antonello A.},
    editor={Jameel, Feroz and Hershenson, Susan and Khan, Mansoor A. and Martin-Moe, Sheryl},
    title={Using Mathematical Modeling and Prior Knowledge for {Q}b{D} in Freeze-Drying Processes},
    booktitle={Quality by Design for Biopharmaceutical Drug Product Development},
    year={2015},
    publisher={Springer},
    address={New York},
    pages={565--593},
    doi = {10.1007/978-1-4939-2316-8_23}
}

@article{Muramatsu2022mRNA,
    title = {Lyophilization provides long-term stability for a lipid nanoparticle-formulated, nucleoside-modified m{RNA} vaccine},
    journal = {Molecular Therapy},
    volume = {30},
    numbers = {5},
    pages = {1941-1951},
    year = {2022},
    doi = {10.1016/j.ymthe.2022.02.001},
    author = {Hiromi Muramatsu and Kieu Lam and Csaba Bajusz and Dorottya Laczkó and Katalin Karikó and Petra Schreiner and Alan Martin and Peter Lutwyche and James Heyes and Norbert Pardi},
}

@article{Meulewaeter2023mRNA,
    title = {Continuous freeze-drying of messenger {RNA} lipid nanoparticles enables storage at higher temperatures},
    journal = {Journal of Controlled Release},
    volume = {357},
    pages = {149-160},
    year = {2023},
    doi = {10.1016/j.jconrel.2023.03.039},
    author = {Sofie Meulewaeter and Gust Nuytten and Miffy H. Y. Cheng and Stefaan C. {De Smedt} and Pieter R. Cullis and Thomas {De Beer} and Ine Lentacker and Rein Verbeke},
}

@article{Fissore2018Review,
    author = {Davide Fissore and Roberto Pisano and Antonello A. Barresi},
    title = {Process analytical technology for monitoring pharmaceuticals freeze-drying – {A} comprehensive review},
    journal = {Drying Technology},
    volume = {36},
    numbers = {15},
    pages = {1839--1865},
    year  = {2018},
    doi = {10.1080/07373937.2018.1440590},
}

@article{Meyer2015SpinFreezing,
    title = {Evaluation of spin freezing versus conventional freezing as part of a continuous pharmaceutical freeze-drying concept for unit doses},
    journal = {International Journal of Pharmaceutics},
    volume = {496},
    numbers = {1},
    pages = {75-85},
    year = {2015},
    doi = {10.1016/j.ijpharm.2015.05.025},
    author = {L. {De Meyer} and P.-J. {Van Bockstal} and J. Corver and C. Vervaet and J. P. Remon and T. {De Beer}},
}

@article{Pisano2019ReviewContLyo,
    title = {Achieving continuous manufacturing in lyophilization: Technologies and approaches},
    journal= {European Journal of Pharmaceutics and Biopharmaceutics},
    volume = {142},
    pages = {265-279},
    year = {2019},
    doi = {10.1016/j.ejpb.2019.06.027},
    author = {Roberto Pisano and Andrea Arsiccio and Luigi C. Capozzi and Bernhardt L. Trout},
}

@article{Capozzi2019ContLyo_SuspendedVials,
    author = {Capozzi, Luigi C. and Trout, Bernhardt L. and Pisano, Roberto},
    title = {From Batch to Continuous: Freeze-Drying of Suspended Vials for Pharmaceuticals in Unit-Doses},
    journal = {Industrial \& Engineering Chemistry Research},
    volume = {58},
    numbers = {4},
    pages = {1635-1649},
    year = {2019},
    doi = {10.1021/acs.iecr.8b02886},
}

@article{Ishwarya2015SprayFD,
    title = {Spray-freeze-drying: A novel process for the drying of foods and bioproducts},
    journal = {Trends in Food Science \& Technology},
    volume = {41},
    numbers = {2},
    pages = {161-181},
    year = {2015},
    issn = {0924-2244},
    doi = {10.1016/j.tifs.2014.10.008},
    author = {S. Padma Ishwarya and C. Anandharamakrishnan and Andrew G. F. Stapley},
}

@article{Hottot2006Freezing,
    author = {Aurélie   Hottot  and  Roman   Peczalski  and  Séverine   Vessot  and  Julien   Andrieu},
    title = {Freeze-Drying of Pharmaceutical Proteins in Vials: Modeling of Freezing and Sublimation Steps},
    journal = {Drying Technology},
    volume = {24},
    numbers = {5},
    pages = {561-570},
    year  = {2006},
    doi ={10.1080/07373930600626388},
}

@article{Nakagawa2007Freezing,
    author = {Nakagawa, Kyuya and Hottot, Aurélie and Vessot, Séverine and Andrieu, Julien},
    title = {Modeling of freezing step during freeze-drying of drugs in vials},
    journal = {AIChE Journal},
    volume = {53},
    numbers = {5},
    pages = {1362-1372},
    doi = {10.1002/aic.11147},
    year = {2007}
}

@article{Arsiccio2017CrystalSize,
    author = {Arsiccio, Andrea and Barresi, Antonello A. and Pisano, Roberto},
    title = {Prediction of Ice Crystal Size Distribution after Freezing of Pharmaceutical Solutions},
    journal = {Crystal Growth \& Design},
    volume = {17},
    numbers = {9},
    pages = {4573-4581},
    year = {2017},
    doi = {10.1021/acs.cgd.7b00319},
}

@article{Deck2022FreezingLumped,
    title = {Stochastic shelf-scale modeling framework for the freezing stage in freeze-drying processes},
    journal = {International Journal of Pharmaceutics},
    volume = {613},
    pages = {121276},
    year = {2022},
    doi = {10.1016/j.ijpharm.2021.121276},
    author = {Leif-Thore Deck and David R. Ochsenbein and Marco Mazzotti},
}

@article{Deck2024Freezing2D,
    title = {Modeling the freezing process of aqueous solutions considering thermal gradients and stochastic ice nucleation},
    journal = {Chemical Engineering Journal},
    volume = {483},
    pages = {148660},
    year = {2024},
    doi = {10.1016/j.cej.2024.148660},
    author = {Leif-Thore Deck and Andraž Košir and Marco Mazzotti},
}

@article{Litchfield1979Model,
    title = {An adsorption-sublimation model for a freeze dryer},
    journal = {Chemical Engineering Science},
    volume = {34},
    numbers = {9},
    pages = {1085-1090},
    year = {1979},
    doi = {10.1016/0009-2509(79)85013-7},
    author = {R. J. Litchfield and A. I. Liapis},
}

@article{Liapis1994Original,
    title = {A theory for the primary and secondary drying stages of the freeze-drying of pharmaceutical crystalline and amorphous solutes: {Comparison} between experimental data and theory},
    journal = {Separations Technology},
    volume = {4},
    numbers = {3},
    pages = {144-155},
    year = {1994},
    doi = {10.1016/0956-9618(94)80017-0},
    author = {A. I. Liapis and R. Bruttini}
}

@article{Sadikoglu1997Modeling,
    author = {H. Sadikoglu and A. I. Liapis },
    title = {Mathematical Modelling of the Primary and Secondary Drying Stages of Bulk Solution Freeze-Drying in Trays: Parameter Estimation and Model Discrimination by Comparison of Theoretical Results With Experimental Data},
    journal = {Drying Technology},
    volume = {15},
    numbers = {3-4},
    pages = {791-810},
    year  = {1997},
    doi = {10.1080/07373939708917262},
}

@article{Sheehan1998Modeling,
    author = {Sheehan, P. and Liapis, A. I.},
    title = {Modeling of the primary and secondary drying stages of the freeze drying of pharmaceutical products in vials: {N}umerical results obtained from the solution of a dynamic and spatially multi-dimensional lyophilization model for different operational policies},
    journal = {Biotechnology and Bioengineering},
    volume = {60},
    numbers = {6},
    pages = {712-728},
    doi = {10.1002/(SICI)1097-0290(19981220)60:6<712::AID-BIT8>3.0.CO;2-4},
    year = {1998}
}

@article{Pikal2005Model,
    author = {M. J. Pikal and W. J. Mascarenhas and H. U. Akay and S. Cardon and Chandan Bhugra and F. Jameel and S. Rambhatla},
    title = {The Nonsteady State Modeling of Freeze Drying: In-Process Product Temperature and Moisture Content Mapping and Pharmaceutical Product Quality Applications},
    journal = {Pharmaceutical Development and Technology},
    volume = {10},
    numbers = {1},
    pages = {17-32},
    year  = {2005},
    doi = {10.1081/PDT-35869},
}

@article{Veraldi2008SimBatchModels,
    title = {Development of simplified models for the freeze-drying process and investigation of the optimal operating conditions},
    author = {Salvatore A. Velardi and Antonello A. Barresi},
    journal = {Chemical Engineering Research and Design},
    volume = {86},
    numbers = {1},
    pages = {9-22},
    year = {2008},
    doi = {10.1016/j.cherd.2007.10.007},
}

@article{Pisano2010control,
    title = {In-Line Optimization and Control of an Industrial Freeze-Drying Process for Pharmaceuticals},
    journal = {Journal of Pharmaceutical Sciences},
    volume = {99},
    numbers = {11},
    pages = {4691-4709},
    year = {2010},
    doi = {10.1002/jps.22166},
    author = {Roberto Pisano and Davide Fissore and Salvatore A. Velardi and Antonello A. Barresi},
}

@article{Chen2015FEMmodel,
    title={Finite Element Method {(FEM)} Modeling of Freeze-drying: Monitoring Pharmaceutical Product Robustness During Lyophilization},
    author={Xiaodong Chen and Vikram Sadineni and Mita Maity and Yong Quan and Matthew Enterline and Rao V. Mantri},
    journal={AAPS PharmSciTech},
    year={2015},
    volume={16},
    pages={1317-1326},
    doi={10.1208/s12249-015-0318-9}
}

@article{Bano2020LumpedDrying,
    author = {Bano, Gabriele and De-Luca, Riccardo and Tomba, Emanuele and Marcelli, Agnese and Bezzo, Fabrizio and Barolo, Massimiliano},
    title = {Primary Drying Optimization in Pharmaceutical Freeze-Drying: A Multivial Stochastic Modeling Framework},
    journal = {Industrial \& Engineering Chemistry Research},
    volume = {59},
    numbers = {11},
    pages = {5056-5071},
    year = {2020},
    doi = {10.1021/acs.iecr.9b06402},
}

@article{Fissore2011SecDryingMonitor,
    title = {Monitoring of the Secondary Drying in Freeze-Drying of Pharmaceuticals},
    journal = {Journal of Pharmaceutical Sciences},
    volume = {100},
    numbers = {2},
    pages = {732-742},
    year = {2011},
    doi = {10.1002/jps.22311},
    author = {Davide Fissore and Roberto Pisano and Antonello A. Barresi}
}

@article{Sahni2017Simplified,
    title = {Modeling the Secondary Drying Stage of Freeze Drying: Development and Validation of an {Excel-based} Model},
    journal = {Journal of Pharmaceutical Sciences},
    volume = {106},
    numbers = {3},
    pages = {779-791},
    year = {2017},
    doi = {10.1016/j.xphs.2016.10.024},
    author = {Ekneet K. Sahni and Michael J. Pikal},
}

@article{Yoon2021Sec0D1D3D,
    title = {Understanding Heat Transfer During the Secondary Drying Stage of Freeze Drying: Current Practice and Knowledge Gaps},
    journal = {Journal of Pharmaceutical Sciences},
    volume = {111},
    numbers = {2},
    pages = {368-381},
    year = {2022},
    doi = {10.1016/j.xphs.2021.09.032},
    author = {Kyu Yoon and Vivek Narsimhan},
}

@article{Bockstal2017ContLyo_ModelPrimary,
    title = {Mechanistic modelling of infrared mediated energy transfer during the primary drying step of a continuous freeze-drying process},
    journal = {European Journal of Pharmaceutics and Biopharmaceutics},
    volume = {114},
    pages = {11-21},
    year = {2017},
    doi = {10.1016/j.ejpb.2017.01.001},
    author = {Pieter-Jan {Van Bockstal} and Séverine Thérèse F. C. Mortier and Laurens {De Meyer} and Jos Corver and Chris Vervaet and Ingmar Nopens and Thomas {De Beer}},
}

@article{Nuytten2021SpinFreezing,
    author = {Nuytten, Gust and Revatta, Susan Ríos and Van Bockstal, Pieter-Jan and Kumar, Ashish and Lammens, Joris and Leys, Laurens and Vanbillemont, Brecht and Corver, Jos and Vervaet, Chris and De Beer, Thomas},
    title = {Development and Application of a Mechanistic Cooling and Freezing Model of the Spin Freezing Step within the Framework of Continuous Freeze-Drying},
    journal = {Pharmaceutics},
    volume = {13},
    year = {2021},
    numbers = {12},
    pages = {2076},
    doi = {10.3390/pharmaceutics13122076},
}

@article{Sebastiao2019SprayFD,
    title = {Bulk Dynamic Spray Freeze-Drying Part 1: Modeling of Droplet Cooling and Phase Change},
    journal = {Journal of Pharmaceutical Sciences},
    volume = {108},
    numbers = {6},
    pages = {2063-2074},
    year = {2019},
    issn = {0022-3549},
    doi = {10.1016/j.xphs.2019.01.009},
    author = {Israel B. Sebastião and Bakul Bhatnagar and Serguei Tchessalov and Satoshi Ohtake and Matthias Plitzko and Bernhard Luy and Alina Alexeenko},
}

@misc{Bruttini1993ContLyoPatent,
    title={Continuous Freeze Drying Apparatus},
    author={Roberto Bruttini},
    note={{U.S.} Patent 5 269 077, Dec 14, 1993}
}

@misc{Gottfried1973ContLyoPatent,
    title={Continuous Freeze Dryer},
    author={Herbert Gottfried},
    note={{U.S.} Patent 3 731 392, May 8, 1973}
}

@book{Mills1995HeatTransfer,
    title={Heat and Mass Transfer},
    author={Mills, A.},
    doi = {10.4324/9780203752173},
    editions = {1st},
    year={1995},
    address = {New York},
    publisher={Routledge}
}

@article{Colucci2020CrystalSize,
    title = {A new mathematical model for monitoring the temporal evolution of the ice crystal size distribution during freezing in pharmaceutical solutions},
    journal = {European Journal of Pharmaceutics and Biopharmaceutics},
    volume = {148},
    pages = {148-159},
    year = {2020},
    issn = {0939-6411},
    doi = {10.1016/j.ejpb.2020.01.004},
    author = {Domenico Colucci and Davide Fissore and Antonello A. Barresi and Richard D. Braatz},
}

@article{Dyer1968HeatTransferLimit,
    title = {The role of convection in drying},
    journal = {Chemical Engineering Science},
    volume = {23},
    numbers = {9},
    pages = {965-970},
    year = {1968},
    issn = {0009-2509},
    doi = {10.1016/0009-2509(68)87082-4},
    author = {D. F. Dyer and J. E. Sunderland},
}

@article{Jafar2003HeatTransferLimit,
    author = {Farial Jafar and Mohammed Farid},
    title = {Analysis of Heat and Mass Transfer in Freeze Drying},
    journal = {Drying Technology},
    volume = {21},
    numbers = {2},
    pages = {249-263},
    year  = {2003},
    publisher = {Taylor & Francis},
    doi = {10.1081/DRT-120017746},
}

@article{Srisuma2023Analytical,
    title = {Analytical solutions for the modeling, optimization, and control of microwave-assisted freeze drying},
    journal = {Computers \& Chemical Engineering},
    volume = {177},
    pages = {108318},
    year = {2023},
    issn = {0098-1354},
    doi = {10.1016/j.compchemeng.2023.108318},
    author = {Prakitr Srisuma and George Barbastathis and Richard D. Braatz},
}

@article{Srisuma2024Obs,
    title = {Real-time estimation of bound water concentration during lyophilization with temperature-based state observers},
    journal = {International Journal of Pharmaceutics},
    volume = {665},
    pages = {124693},
    year = {2024},
    doi = {10.1016/j.ijpharm.2024.124693},
    author = {Prakitr Srisuma and George Barbastathis and Richard D. Braatz},
}

@article{Pikal1990SecDrying_Kinetics,
    title = {The secondary drying stage of freeze drying: drying kinetics as a function of temperature and chamber pressure},
    journal = {International Journal of Pharmaceutics},
    volume = {60},
    numbers = {3},
    pages = {203-207},
    year = {1990},
    issn = {0378-5173},
    doi = {10.1016/0378-5173(90)90074-E},
    author = {M. J. Pikal and S. Shah and M. L. Roy and R. Putman},
}

@article{Kramer2002VISF,
    title = {Freeze‐drying using vacuum‐induced surface freezing},
    journal = {Journal of Pharmaceutical Sciences},
    volume = {91},
    numbers = {2},
    pages = {433-443},
    year = {2002},
    doi = {10.1002/jps.10035},
    author = {Martin Kramer and Bernd Sennhenn and Geoffrey Lee},
}

@article{Srisuma2024Rad,
    title = {{Mechanistic modeling and analysis of thermal radiation in conventional, microwave-assisted, and hybrid freeze drying for biopharmaceutical manufacturing}},
    year = {2024},
    journal = {International Journal of Heat and Mass Transfer},
    author = {Srisuma, Prakitr and Barbastathis, George and Braatz, Richard D.},
    volume = {221},
    doi = {10.1016/j.ijheatmasstransfer.2023.125023},
    pages = {125023},
}

@article{Gan2005WallTemp,
    title = {Freeze-drying of pharmaceuticals in vials on trays: effects of drying chamber wall temperature and tray side on lyophilization performance},
    journal = {International Journal of Heat and Mass Transfer},
    volume = {48},
    numbers = {9},
    pages = {1675-1687},
    year = {2005},
    doi = {10.1016/j.ijheatmasstransfer.2004.12.004},
    author = {K. H. Gan and R. Bruttini and O. K. Crosser and A. I. Liapis},
}

@book{Smith2018Thermo,
    title={Introduction to Chemical Engineering Thermodynamics},
    author={J. M. Smith and H. C. {Van Ness} and M. M. Abbott and M. T. Swihart},
    edition = {8th},
    year={2018},
    address = {New York},
    publisher={McGraw-Hill Education}
}

@article{Glatz2018Nucleation,
    author = {Glatz, Brittany and Sarupria, Sapna},
    title = {Heterogeneous Ice Nucleation: Interplay of Surface Properties and Their Impact on Water Orientations},
    journal = {Langmuir},
    volume = {34},
    numbers = {3},
    pages = {1190-1198},
    year = {2018},
    doi = {10.1021/acs.langmuir.7b02859},
}

@article{Sircar2000LDF,
    author={Sircar, S. and Hufton, J. R.},
    title={Why Does the Linear Driving Force Model for Adsorption Kinetics Work?},
    journal={Adsorption},
    year={2000},
    volume={6},
    pages={137-147},
    doi={10.1023/A:1008965317983},
}

@book{Chang2010Chem,
    title={Chemistry},
    author={Raymond Chang},
    edition = {10th},
    year={2010},
    address = {New York},
    publisher={McGraw-Hill}
}

@book{Incropera2007HeatTransfer,
    title={Fundamentals of Heat and Mass Transfer},
    author={Frank P. Incropera and David P. Dewitt and Theodore L. Bergman and Adrienne S. Lavine},
    edition = {6th},
    year={2007},
    publisher={John Wiley \& Sons},
    address = {New Jersey}
}

@book{Bird2002Transport,
    title={Transport Phenomena},
    author={Bird, R. B. and Lightfoot, E. N. and Stewart, W. E.},
    address = {New York},
    edition = {2nd},
    year={2002},
    publisher={John Wiley \& Sons}
}

@book{Crank1979Diffusion,
    title={The Mathematics of Diffusion},
    author={Crank, J.},
    edition = {2nd},
    publisher={Oxford University Press},
    address ={London},
    year={1975}
}

@book{Schiesser1991MOL,
    title={The Numerical Method of Lines: Integration of Partial Differential Equations},
    author={Schiesser, W. E.},
    year={1991},
    publisher={Academic Press},
    address = {San Diego}
}

@article{Pisano2017VISF,
title = {Prediction of product morphology of lyophilized drugs in the case of {Vacuum Induced Surface Freezing}},
journal = {Chemical Engineering Research and Design},
volume = {125},
pages = {119-129},
year = {2017},
doi = {10.1016/j.cherd.2017.07.004},
author = {Roberto Pisano and Luigi C. Capozzi},
}

@article{Pisano2025Freezing,
    author = {Vincenzo Massotti and Fiora Artusio and Antonello A. Barresi and Roberto Pisano},
    title = {Mathematical modelling of thermal interactions during freezing: Effects on product morphology and drying behaviour},
    journal = {European Journal of Pharmaceutical Sciences},
    volume = {210},
    year = {2025},
    pages = {107112},
    doi = {10.1016/j.ejps.2025.107112}
}

@article{Stratta2024Stopper,
    author = {Lorenzo Stratta and Roberto Pisano},
    title = {The impact of the stopper position and geometry on the freeze-drying cycle of pharmaceutical products},
    journal = {Drying Technology},
    volume = {42},
    pages = {1999--2011},
    year = {2024},
    publisher = {Taylor \& Francis},
    doi = {10.1080/07373937.2024.2399295},
}


\clearpage
\begin{figure}
\textbf{Table of Contents}\\
  \includegraphics{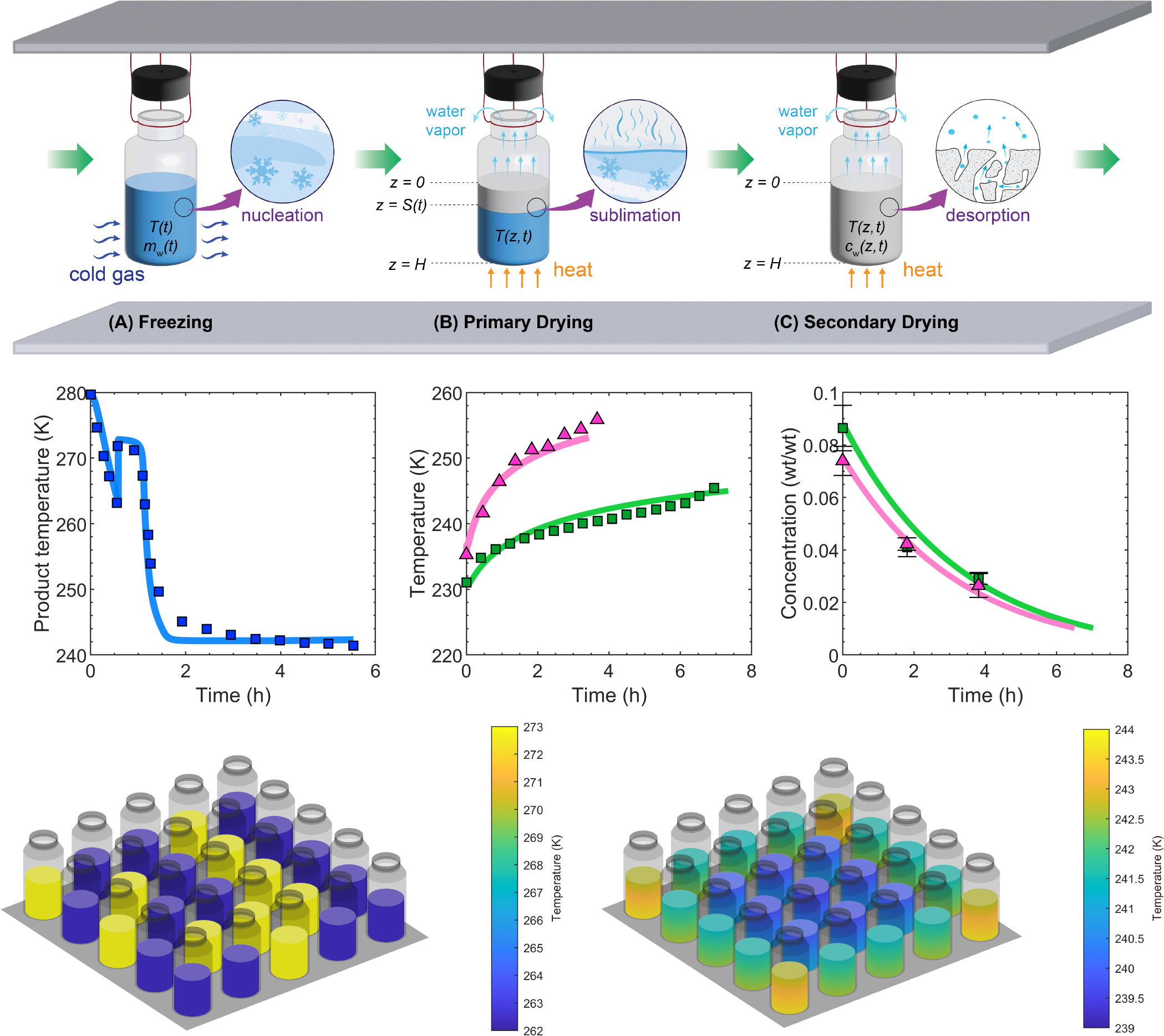}
  \medskip
  \caption*{This article presents the first mechanistic model for continuous lyophilization (aka freeze drying). The model accurately describes the key phenomena and critical process parameters in continuous lyophilization, which can be used for process design, optimization, and control. Results from this work could drive the transition of (bio)pharmaceutical freeze drying from batch processes to continuous manufacturing.}
\end{figure}

\end{document}